\documentstyle[emulateapj,apjfonts,epsf]{article}

\slugcomment{Accepted by The Astrophysical Journal, November 11, 1999}

\begin{document}

\title{Observations of 20 millisecond pulsars in 47~Tucanae at 20\,cm}

\author{F. Camilo,\altaffilmark{1,2}
  D. R. Lorimer,\altaffilmark{3,4}
  P. Freire,\altaffilmark{1}
  A. G. Lyne,\altaffilmark{1} \and
  R. N. Manchester\altaffilmark{5}}
\medskip
\affil{\altaffilmark{1}University~of~Manchester,~Jodrell~Bank~Observatory,~Macclesfield,~Cheshire,~SK11~9DL,~UK}
\affil{\altaffilmark{3}Max Planck Institut f\"ur Radioastronomie, Auf
dem H\"ugel 69, 53121, Bonn, Germany}
\affil{\altaffilmark{5}Australia Telescope National Facility, CSIRO,
  P.O.~Box~76, Epping, NSW~1710, Australia}
\altaffiltext{2}{Marie Curie Fellow.  Present address: Columbia
  Astrophysics Laboratory, 550 West 120th Street, New York, NY~10027.
  E-mail: fernando@astro.columbia.edu}
\altaffiltext{4}{Present address: NAIC, Arecibo Observatory, HC3 Box
  53995, Arecibo, PR~00613}

\bigskip

\begin{abstract}
We have used a new observing system on the Parkes radio telescope to
carry out a series of pulsar observations of the globular cluster
47~Tucanae at 20-cm wavelength.  We detected all 11 previously known
pulsars, and have discovered nine others, all of which are millisecond
pulsars in binary systems.  We have searched the data for relatively
short orbital period systems, and found one pulsar with an orbital
period of 96\,min, the shortest of any known radio pulsar.

The increased rate of detections with the new system resulted in
improved estimates of the flux density of the previously known pulsars,
determination of the orbital parameters of one of them, and a coherent
timing solution for another one.  Five of the pulsars now known in
47~Tucanae have orbital periods of a few hours and implied companion
masses of only $\sim 0.03$\,M$_\odot$.  Two of these are eclipsed at
some orbital phases, while three are seen at all phases at 20\,cm but
not always at lower frequencies.  Four and possibly six of the other
binary systems have longer orbital periods and companion masses $\sim
0.2$\,M$_\odot$, with at least two of them having relatively large
orbital eccentricities.  All 20 pulsars have rotation periods in the
range 2--8\,ms.

\end{abstract}

\keywords{binaries: eclipsing --- binaries: general --- globular
clusters: individual (47~Tucanae) --- pulsars: general}

\section{Introduction}\label{sec:intro}

Soon after the discovery of the first millisecond pulsar
(\cite{bkh+82}), Alpar et~al.~(1982)\nocite{acrs82} proposed that
low-mass X-ray binaries (LMXBs) are the progenitors of millisecond
pulsars.  Given the high incidence of LMXBs in globular clusters,
searching for millisecond pulsars in clusters became a natural goal.
After much effort, the first such object was discovered, an isolated
pulsar with a period of 3\,ms in M28 (\cite{lbm+87}).  Following
searches of many clusters in the late 1980s and early 1990s, over 30
pulsars are now known in 13 clusters.  In contrast to most pulsars in
the Galactic disk, the majority of cluster pulsars have short rotation
periods and are often members of binary systems (see \cite{ka96} for a
review).

Searches for millisecond pulsars in globular clusters encounter two
main difficulties.  Firstly, since globular clusters are on average
more distant than the $\sim 2$\,kpc within which most large-area
surveys have been sensitive to millisecond pulsars (see, e.g.,
\cite{cnt96}), much longer observation times are typically required
than for Galactic searches.  This in turn increases the complexity of
the data-reduction task.  Such observation times are feasible only
because of the small number of search targets.  Secondly, for binary
pulsars there is the added difficulty that the apparent pulse period
may change significantly as the pulsar experiences orbital
accelerations during the longer integration.  If this issue is not
addressed at the data reduction stage --- and with significant
exceptions (\cite{agk+90}) it rarely is, because of the massive
computational requirements --- many relatively bright binary pulsars
may remain undetected.

Despite these difficulties, the rewards of a thorough search can be
significant.  Because of the high probability of close stellar
interactions, unusual binary systems not otherwise expected in the
Galactic disk may form in clusters.  In addition, for clusters where
several pulsars are found and studied with the detail allowed by
high-precision timing measurements, much can be learned about the
clusters themselves.  Applications to date include a determination of
the core mass distribution from measurements of pulsar accelerations
and positions, and limits on intra-cluster medium properties, by
measuring the dispersion of various pulsars (\cite{phi92b};
\cite{and92}).

\begin{deluxetable}{lll}
\scriptsize
\tablecaption{\label{tab:47tuc}Observed and derived properties of 47~Tucanae.}
\tablecolumns{3}
\tablehead{
\colhead{Parameter} & 
\colhead{Value}     & 
\colhead{Reference} \nl}
\startdata
Center R.A., $\alpha_{\rm 47~Tuc}$  (J2000) \dotfill & $00^{\rm h} 24^{\rm m} 05\fs9$ &\cite{gysb92}\nl
Center Decl., $\delta_{\rm 47~Tuc}$ (J2000) \dotfill & $-72^\circ04'51\farcs1$        &\cite{gysb92}\nl
Distance             \dotfill & 4.6\,kpc                       &\cite{web85} \nl
Central mass density, $\rho(0)$ \dotfill & $4\times10^5\,$M$_\odot\,{\rm pc}^{-3}$ & 
                                               This work\tablenotemark{\star}\nl
Angular tidal radius\tablenotemark{\dagger} \dotfill & $40'$ (54\,pc)  &\cite{dac79} \nl
Angular core radius\tablenotemark{\ddagger} \dotfill & $12\farcs2$ (0.27\,pc) & \cite{dps+96} \nl
Escape velocity     \dotfill  & 56.8\,km\,s$^{-1}$             &\cite{web85} \nl
Velocity dispersion \dotfill  & 13.2\,km\,s$^{-1}$             &\cite{web85} \nl
\enddata

\tablenotetext{\star}{The value quoted is a revision of Webbink's
(1985) catalog entry (see \S\ref{sec:density} for details).}

\tablenotetext{\dagger}{Also known as the ``limiting radius'', this is
the radius at which the King model surface brightness--radius profile
drops to zero (\cite{kin62}).}

\tablenotetext{\ddagger}{The core radius is a scale factor used in the
King model, and denotes the radius at which the number density of stars
in the cluster per unit area is half the central value.}

\end{deluxetable}

The basic observed and derived properties of 47~Tucanae (NGC~104) used
throughout this paper are summarized in Table~\ref{tab:47tuc}.  The
fact that it is a relatively nearby cluster with a high stellar density
makes it an attractive target for pulsar searches.  47~Tuc together
with M15 have been by far the most prolific cluster targets for pulsar
searches to date, with surveys revealing 11 and eight pulsars
respectively (\cite{mlr+91}; \cite{rlm+95}; \cite{and92}).  The
observed population of pulsars in these two clusters is very different:
while those in 47~Tuc are all millisecond pulsars, the pulsars in M15
are more varied in their properties, including slowly-spinning pulsars
and a rare double neutron-star binary (\cite{pakw91}).  It is also
worth noting that although all pulsars in M15 have had positions and
period derivatives determined using pulsar timing techniques, this has
been possible to date for only two of the pulsars in 47~Tuc.  This is
due mainly to the different telescopes used and different
interstellar scintillation characteristics: at Arecibo, Puerto Rico,
the majority of the distant pulsars in M15, whose apparent fluxes vary
little, are detectable on most days, while at Parkes, Australia, most
pulsars in 47~Tuc are visible only when the interstellar weather is
propitious and the pulsar signals are amplified in the observing radio
band by scintillation.  Unfortunately this has meant that many of the
scientific objectives suggested by the wealth of pulsars known in
47~Tuc remain unfulfilled.

In mid-1997 a new, sensitive system for observing pulsars at a
wavelength of 20\,cm became available at the Parkes telescope.  Pulsars
are in general steep-spectrum sources (\cite{lylg95}), and all pulsars
in 47~Tuc were discovered at a wavelength of 50 or 70\,cm (Manchester
et~al.~1990, 1991; \cite{rlm+95}).\nocite{mld+90,mlr+91}  However, the
new system provided such performance improvements that observations to
study the known pulsars and to find previously undetected pulsars
seemed desirable.  These observations were so successful in the rate of
detection of previously known pulsars and with discovery of new ones,
that we have moved to observing the cluster almost exclusively with the
new system.  In this paper we report on the discovery of nine binary
millisecond pulsars in 47~Tuc and give improved parameters for some of
the previously known pulsars.

In \S\ref{sec:data} we describe the observing system used, list the
observations made, and outline the data reduction carried out.  In
\S\ref{sec:sens} we estimate the sensitivity of the survey for isolated
and long orbital period pulsars, and compare it to that of past surveys
of 47~Tuc.  In \S\ref{sec:res} we summarize the results, including
parameters for the newly discovered pulsars and updated parameters for
some previously known systems.  Finally, in \S\ref{sec:disc}, we
discuss some limitations, and implications, of our survey and its
results.

\section{Observations and Data Analysis}\label{sec:data}

Since 1997 October we have been using the Parkes telescope together
with the pulsar multibeam data-acquisition system to observe the
globular cluster 47~Tuc at a central radio frequency of 1374\,MHz.  The
system consists of a cooled, dual linear polarization, 13-beam receiver
package (\cite{swb+96}), and new filter banks, single-bit digitizer,
and data-acquisition computer.

Each of the 13 beams has a half-power diameter of about $14'$ on the
sky.  Since a single beam encompasses the great majority of the cluster
stars (cf. Table~\ref{tab:47tuc}), we record data from the central beam
only (but see discussion in \S\ref{sec:selection}), while pointing the
telescope at 47~Tuc~C, the first pulsar discovered in this cluster.
Total-power signals from each of 96 contiguous frequency channels, each
3\,MHz wide, are one-bit sampled every 125\,$\mu$s and, together with
relevant telescope information, are written to magnetic tape for
off-line processing.  Data are usually collected on days when a pulsar
survey of the Galactic plane is also underway (\cite{lcm+00}), and are
recorded from 47~Tuc for up to five hours daily.  Table~\ref{tab:obs}
contains a log of all observations reported in this paper.

\begin{deluxetable}{llllllllllllllll}
\tiny
\tablecaption{\label{tab:obs}Log of observations and detections.}
\tablecolumns{16}
\renewcommand{\tabcolsep}{1.0mm}
\tablehead{
\colhead{Date}    & 
\colhead{Number of}  & 
\colhead{C}       & 
\colhead{D}       & 
\colhead{E}       & 
\colhead{F}       & 
\colhead{G}       & 
\colhead{H}       & 
\colhead{I}       & 
\colhead{J}       & 
\colhead{L}       & 
\colhead{M}       & 
\colhead{N}       & 
\colhead{O}       & 
\colhead{Q}       & 
\colhead{U}       \nl
\colhead{(MJD)}       &
\colhead{integrations} &
\colhead{}         &
\colhead{}         &
\colhead{}         &
\colhead{}         &
\colhead{}         &
\colhead{}         &
\colhead{}         &
\colhead{}         &
\colhead{}         &
\colhead{}         &
\colhead{}         &
\colhead{}         &
\colhead{}         &
\colhead{}         \nl}
\startdata
50683 & 16 &   0/4.6 &   2/7.0 & 14/19.8 &   0/2.3 &   0/2.5 & 10/17.0 &   0/2.4 &   6/8.8 &         &   0/2.5 &         &   0/2.9 &         &        \nl 
50686 & 16 &   4/7.6 &   0/3.2 &   0/4.4 & 16/33.5 &   0/3.4 &         &         & 16/12.2 &   0/3.1 &         &         &  4/10.1 &         &        \nl 
50689 & 16 &   5/9.8 &         &   0/3.1 &   0/3.4 &   0/2.9 &         &   3/6.0 & 11/12.9 &   0/2.8 &         & 16/18.2 &   0/2.7 &         &        \nl 
50690 & 16 & 16/21.9 &   0/2.8 &   0/3.2 &   0/6.1 &   0/3.3 &         &         &   1/6.6 &         &         &         &   3/8.6 &   0/2.0 &        \nl 
50739 & 5  &  5/20.9 &  3/10.2 &         &         &         &         &   0/7.4 &  5/31.5 &         &         &         &         &         &        \nl 
50740 & 8  &   0/4.4 &   1/7.2 &   0/7.2 &         &   0/3.7 &         &   2/8.8 &  8/13.4 &         &   0/3.8 &         &   0/5.3 &  7/13.5 &  0/4.2 \nl 
50741 & 8  &         &  8/11.9 &  8/19.7 &         &         &         &   2/9.8 &  8/46.5 &         &         &         &   0/4.1 &   0/4.5 &  0/3.0 \nl 
50742 & 8  &  8/15.7 &   0/4.1 &         &   2/6.5 &   0/3.1 &  8/16.6 &   0/6.1 &  6/12.6 &   0/4.0 &         &         &         &         &  0/4.0 \nl 
50743 & 3  &   0/7.9 &         &         &  3/21.9 &  3/12.8 &         &         &  3/17.8 &         &         &         &         &         &        \nl
50744 & 8  &  8/15.3 &         &   0/3.2 &   0/3.6 &  8/14.7 &   0/4.2 &   0/2.9 &  4/11.1 &         &         &         &   0/3.2 &         &        \nl 
50745 & 8  &   0/2.8 &  8/22.6 &         &   0/5.4 &         &  5/11.3 &   0/3.7 &  8/26.7 &         &         &         &   0/3.5 &         &        \nl 
50746 & 8  &   0/5.4 &  8/17.7 &         &   0/3.1 &         &   0/6.5 &   0/3.6 &  8/32.2 &         &         &         &         &   0/6.0 &        \nl 
50748 & 4  &  4/10.5 &   0/3.7 &         &         &   0/3.9 &         &         &         &         &  4/14.5 &         &   0/4.3 &         &        \nl 
50980 & 8  &  8/22.5 &  8/33.4 &  8/20.7 &   1/4.9 &         &         &         &   0/3.8 &         &         &   0/3.4 &         &   0/2.7 &  1/8.4 \nl 
50981 & 16 & 16/63.2 &   5/9.3 &         & 15/18.8 &   1/4.2 &         &         &   7/7.9 &         &   1/6.3 &         &   0/3.0 &         &        \nl 
50982 & 16 &   2/5.8 & 16/22.7 &  9/13.3 &   1/5.7 &         &   0/2.5 &         &   3/5.4 &         &   1/8.5 &         &   0/4.3 &         &        \nl 
50983 & 4  &   0/5.0 &         &  3/12.6 &   0/6.3 &         &         &         &         &         &         &         &   0/7.4 &         &  0/5.6 \nl 
50992 & 8  &   0/5.4 &   0/6.7 &         &  6/12.9 &         &         &         &   0/4.7 &         &         &         &         &         &  0/3.3 \nl 
50993 & 8  &  8/40.2 &   0/7.2 &   0/5.4 &   0/2.9 &         &         &   1/9.1 &  8/22.8 &  7/11.8 &         &         &         &         &  0/3.0 \nl 
50998 & 16 &   0/2.1 &   3/6.6 &   1/5.4 &   2/5.4 &         &         &         &   0/2.7 &   0/2.7 &         & 14/11.8 &         &         &        \nl 
50999 & 5  &   0/6.4 &  5/14.6 &         &  4/10.8 &         &         &   0/4.1 &   0/6.8 &         &         &         &   0/4.8 &         &  0/3.7 \nl 
51000 & 16 &   0/6.7 &   0/2.1 &         &         &         &         &   0/2.7 &   2/6.7 &   0/3.1 &         &   0/2.3 &   0/2.6 &         &  0/2.2 \nl 
51001 & 10 &  5/10.2 &   0/3.2 &         &         &         &   0/6.0 &         &   0/3.1 &         &         &         &         &   1/2.9 & 4/10.9 \nl 
51002 & 16 &         &   2/5.5 &   1/7.4 &         &         &         &   0/2.8 &16/146.3 &         &   0/2.2 &         &   3/6.4 &         &  3/7.7 \nl 
51003 & 16 &         &   4/8.6 &   4/8.8 &   0/6.1 &   0/2.8 &   0/3.2 &   0/2.6 & 16/36.1 &         &         &         &   0/4.9 &         &  0/5.3 \nl 
51004 & 16 &   0/5.8 &         &   0/5.7 & 12/12.6 &         &         &         & 13/16.2 &         &         &         &         & 10/10.2 &  0/3.0 \nl 
51005 & 12 &   2/7.7 &   0/3.0 &         &   0/2.5 &         &         &         &  9/12.8 &   0/2.7 &         &         &   2/4.4 &         &        \nl
51007 & 10 &   0/4.3 &   0/5.2 &         &         &         &         &         & 10/23.5 &         &         &         &         &   0/2.9 &        \nl 
51012 & 8  &  5/11.4 &         &   0/7.1 &  4/10.9 &   1/6.2 &         &         & 8/186.1 &         &         &         &   3/7.2 &         &  0/4.7 \nl 
51026 & 16 & 16/25.1 &   0/4.4 &   0/4.0 &   0/3.0 &   0/3.6 &         &         &   4/8.3 &         &   0/2.2 &         &   0/3.2 &         &  0/2.4 \nl 
51027 & 7  &  6/15.7 &  7/36.0 &   3/9.9 &         &         &         &   0/7.2 &  7/14.3 &         &   0/3.4 &         &         &         &  0/3.1 \nl 
51029 & 5  &   0/4.4 &         &   1/9.5 &   0/5.4 &         &         &   0/4.1 &  5/26.2 &         &         &         &         &         &        \nl 
51031 & 6  &   0/3.1 &         &   0/4.7 &  4/10.2 &         &         &   0/6.7 &   3/8.0 &         &         &         &   0/5.1 &         &  0/7.1 \nl 
51033 & 6  &         &  3/12.8 &         &         &   0/5.5 &         &         &   0/3.4 &         &         &         &         &  6/20.7 &        \nl 
51036 & 7  &   0/3.5 &  7/15.2 &         &   4/9.9 &         &         &         &   0/4.5 &         &         &         &   0/5.8 &   2/9.6 &  0/3.9 \nl 
51037 & 16 &         & 16/34.1 &         &         &         &   0/2.6 &   0/4.6 & 16/29.7 &         &   0/3.8 &         &         &         &  0/2.2 \nl 
51038 & 6  &  6/18.3 &  5/10.2 &         &   1/8.1 &         &         &   0/3.8 &         &         &         &         &   0/6.2 & 10/10.7 &        \nl 
51039 & 3  &  3/17.2 &  3/19.9 &   1/8.6 &         &         &   0/8.4 &  1/10.3 &         &         &         &         &         &         &        \nl 
51040 & 4  &  4/18.0 &  4/16.8 &         &         &  2/12.1 &         &   0/8.3 &  4/42.0 &         &         &         &   0/5.0 &         &  0/8.2 \nl 
51085 & 16 &   5/8.9 &  10/9.6 &   0/2.8 &   0/3.0 &         &         &   0/2.1 & 16/16.5 &         &   0/3.2 &   0/3.2 &   0/4.4 &         &  0/3.4 \nl 
51086 & 16 & 16/27.9 &   0/2.9 &         &         &   0/2.2 & 15/17.4 &         & 16/44.5 &   0/5.0 &         &         &   1/6.2 &         &  0/2.3 \nl 
51088 & 19 &   0/2.2 &   0/2.9 &         &   0/2.8 &   0/1.9 &   1/5.7 &   0/1.9 &  8/13.0 &         &   0/2.2 &         &   0/2.7 &         &        \nl 
51089 & 4  &  4/13.6 &         &         &   0/4.8 &         &         &  3/10.3 &   0/5.1 &         &   0/4.6 &         &         &         &        \nl 
51090 & 4  &  4/23.2 &  4/20.5 &         &   0/5.7 &         &         &         &  4/16.7 &         &         &         &         &         &  0/7.9 \nl 
51091 & 4  &  4/49.3 &  4/21.3 &         &         &         &  4/21.2 &         &   1/9.3 &         &         &         &         &         &        \nl 
51092 & 8  &  8/29.4 &  7/15.5 &   1/7.7 &  8/29.3 &         &         &  5/13.4 &   0/4.8 &         &   0/6.4 &         &         &   0/2.8 &  0/4.5 \nl 
51093 & 8  &   0/6.8 &   0/3.5 &         &   0/3.2 &         &         &   0/4.6 &         &         &         &         &   3/6.7 &         &        \nl 
51095 & 4  &  4/18.0 &         &   0/6.8 &         &         &         &  3/11.8 &  1/10.7 &         &         &         &         &         &        \nl 
51096 & 5  &  4/30.5 &  5/16.4 &         &   0/4.5 &         &         &         &   0/4.8 &         &         &         &         &         &        \nl 
51097 & 12 &   3/9.8 &  4/12.2 &         &  7/18.2 &         &   1/5.2 &         & 12/88.3 &         &  8/11.2 &         &   0/3.7 &         &        \nl 
51098 & 16 & 15/17.8 &   0/2.4 &   0/3.7 &         &         &         &         &   0/2.9 &   2/6.6 &   2/6.2 &         &         &         &        \nl 
51099 & 19 &   0/2.7 &   4/6.9 & 18/28.6 &   1/3.8 &         &         &         &   0/3.6 &         &         &         &   0/2.6 &         & 10/8.2 \nl 
51100 & 16 &   0/3.5 &         &   0/2.4 &         &   0/3.5 &   1/6.0 &         & 16/16.5 &         &         &         &         &         &  0/6.3 \nl 
51101 & 17 & 16/23.5 &   0/2.3 &         &         &         &         &         &  10/8.7 &         &         &         &         &         &  0/2.8 \nl 
51145 & 16 &   0/3.0 & 10/14.8 & 16/12.6 &   ?/9.0 &         &         &         & 16/18.1 &         &         &   0/2.3 &         &         &        \nl 
51146 & 16 &   3/6.8 &   0/4.9 &         &   0/3.4 &         &         &         &   2/6.6 &         &         &         &         &         &        \nl 
51147 & 8  &   0/3.5 &         &   0/5.5 &         &         &         &         &         &         &         &         &         &         & 6/11.6 \nl 
51148 & 5  &   1/3.6 &  5/31.2 &         &         &         &         &         &  0/13.4 &         &         &         &         &         &  0/6.4 \nl 
51149 & 16 &   0/5.4 &   0/5.1 &         &         &   0/2.4 &         &   0/2.6 & 13/11.4 &   0/2.9 &   0/2.1 &         &   0/2.3 &   0/2.5 &  1/5.2 \nl 
51150 & 16 &   0/5.4 &         &   0/2.8 &         &         &         &   0/2.2 & 14/22.7 &   0/2.3 &         &         &   0/4.0 &         &        \nl 
51151 & 9  &   2/7.7 &  9/15.1 &         &         &         &         &         &   2/5.3 &         &         &         &         &         &        \nl 
51152 & 17 &   1/6.2 &   4/7.6 &         &   0/2.8 &         &         &   0/2.5 & 16/14.5 &         &         &         &   0/4.6 &         &  3/7.3 \nl 
51153 & 16 &   0/3.0 &   0/4.0 &   0/2.7 &         &   0/5.4 &   1/4.8 &         & 15/16.1 &         &         &   0/2.0 &         &         &        \nl 
51154 & 19 & 19/69.6 &   5/9.0 &         &         &         &   6/8.9 &   0/2.3 & 12/14.5 &         &         &         &   0/2.7 &         &        \nl 
51155 & 16 & 14/18.1 &   0/2.1 &   0/4.8 &         &         &   0/2.5 &   0/2.9 &  11/9.9 &         &         &   0/6.2 &   1/6.0 &   0/2.2 &        \nl 
51156 & 16 & 10/49.0 &   3/6.1 &  4/14.4 & 16/32.0 &         &         &         & 16/52.2 &         &         &         &         &         &        \nl 
51157 & 6  &   0/6.9 &  6/23.2 &         &         &         &         &   0/3.2 &  6/53.2 &         &         &         &         &         &        \nl 
51211 & 16 &   0/5.5 &   2/6.4 & 15/20.3 &   4/6.4 &   0/2.0 &   0/4.6 &   0/3.2 &   0/2.2 &   0/2.2 &         &         &   1/4.1 &         &  0/2.7 \nl 
51212 & 16 &   3/5.6 &   0/3.8 &         &   0/2.1 &   0/2.9 &         &   0/2.1 & 16/38.6 &         &         &   0/2.7 &   0/2.3 &   0/2.8 &        \nl 
51213 & 16 &         &   4/6.0 &   0/2.6 &   0/2.1 &   0/7.1 &         &   0/2.8 &   0/2.8 &         &         &         & 15/13.7 &         &13/14.1 \nl 
51214 & 16 &   2/5.9 &   0/2.5 &         &   0/2.9 &         &   0/2.1 &   0/3.7 & 16/18.5 &         &         &         &   0/3.3 &         &        \nl 
51215 & 16 &   3/6.2 &   0/2.1 &   0/2.5 & 15/13.2 &   0/2.1 &   0/3.0 &   0/2.1 &   2/5.4 &         &         &         & 11/12.4 &         &        \nl 
51216 & 16 &   9/9.7 &   1/6.1 & 10/10.0 &   9/9.4 &         &         &   0/2.8 &  9/10.4 &         &         &         & 15/19.7 &         &        \nl 
51217 & 16 &   0/4.0 &   0/4.3 &   0/3.6 &         &   0/2.7 &   7/8.4 &   0/2.8 & 14/28.8 &         &         &         &   0/3.1 &         &  0/2.9 \nl 
51218 & 16 &   2/8.6 &   0/2.6 &   0/3.9 &   0/5.0 &         &         &   0/2.8 & 16/24.1 &         &         &         &   0/4.7 &         &  4/7.0 \nl 

\enddata

\tablecomments{For each observing date we list the number of 17.5\,min
integrations that were independently searched.  For each pulsar on each
day we list the number of integrations in which the pulsar was detected
with the search code according to the criteria summarized in
\S\ref{sec:res}; and the signal-to-noise ratio ($\sigma$) scaled to a 17.5\,min integration,
obtained from summing the entire day's data set using the best
pulsar ephemeris available (e.g., on MJD~51004, 47~Tuc~C was not
detected in any of 16 integrations with $\sigma \ge 9$, but all data
added together using the pulsar's ephemeris yielded $\sigma = 23.2$, or
an average of $\sigma = 5.8$ per integration.  Hence the 0/5.8
notation.  For 47~Tuc~D, even adding all data did not yield an
unambiguous detection --- one with $\sigma \ga 7$ --- and so the
respective entry is blank).}

\end{deluxetable}

The largest continuous data sets recorded are 4.66\,h long,
corresponding to $2^{27}$ time samples.  With the computing resources
presently at our disposal, it is impracticable to reduce such data sets
in single, coherent, Fourier analyses.  Furthermore, since about one
third of the pulsars already known in 47~Tuc are in binary systems,
some with orbital periods of under 3\,h (\cite{rlm+95}), we would
select against such systems if we analyzed each large data set as a
whole.  As a compromise between computational feasibility and
sensitivity (both to weak radio sources and relatively short-period
binaries), we restrict the data analysis reported in this paper to
contiguous blocks of $2^{23}$ time samples (17.5\,min).

The data were de-dispersed to a reference dispersion measure (DM) of
24.6\,cm$^{-3}$\,pc, the DM of 47~Tuc~C.  This was done by delaying
samples from the higher-frequency channels to account for the faster
propagation through the ionized interstellar medium, before summing
data from all channels into a one-dimensional time series.  For the
present analysis, where we are particularly interested in looking for
binary pulsars with short orbital periods, this choice was made largely
to keep the total computational effort manageable.  The pulsars
presently known in 47~Tuc exhibit a range of DMs between 24.1 and
24.7\,cm$^{-3}$\,pc (Table~\ref{tab:spin}).  In the current analysis,
any pulsar with a true DM that is within one unit of
24.6\,cm$^{-3}$\,pc will have its pulse smeared by at most 1\,ms as a
result of this single-DM approximation.

\begin{deluxetable}{llccc}
\scriptsize
\tablecaption{\label{tab:spin}Pulse periods, dispersion measures,
duty-cycles, and flux densities at 20\,cm.}
\tablecolumns{5}
\tablehead{
\colhead{Pulsar}   & 
\colhead{Period}   & 
\colhead{DM}       & 
\colhead{$w_{50}$} &
\colhead{Flux density} \nl
\colhead{}         &
\colhead{(ms)}     &
\colhead{(cm$^{-3}$\,pc)} &
\colhead{(\%)}     &
\colhead{(mJy)}    \nl}
\startdata
\multicolumn{5}{c@{\vspace{1mm}}}{Previously Known} \nl
\tableline
47~Tuc~C & 5.756780 & 24.6 & 13 & 0.36(4) \nl
47~Tuc~D & 5.357573 & 24.7 & 10 & 0.22(3) \nl
47~Tuc~E & 3.536329\tablenotemark{\star} & 24.3 & 25 & 0.21(3) \nl
47~Tuc~F & 2.623579 & 24.4 & 14 & 0.15(2) \nl
47~Tuc~G & 4.040379 & 24.4 & 13 & 0.05(2) \nl
47~Tuc~H & 3.210341\tablenotemark{\star} & 24.4 & 13 & 0.09(2) \nl
47~Tuc~I & 3.484992\tablenotemark{\star} & 24.5 & 14 & 0.09(1) \nl
47~Tuc~J & 2.100634\tablenotemark{\star} & 24.6 & 15 & 0.54(6) \nl
47~Tuc~L & 4.346168 & 24.4 & 15 & 0.04(1) \nl
47~Tuc~M & 3.676643 & 24.4 & 25 & 0.07(2) \nl
47~Tuc~N & 3.053954 & 24.6 & 11 & 0.03(1) \nl
\tableline
\multicolumn{5}{c@{\vspace{-1mm}}}{} \nl
\multicolumn{5}{c@{\vspace{1mm}}}{Newly Discovered} \nl
\tableline
47~Tuc~O & 2.643343\tablenotemark{\star} & 24.4 & 12 & 0.10(1) \nl
47~Tuc~P & 3.643021\tablenotemark{\star} & 24.3 & 15 & \nodata \nl
47~Tuc~Q & 4.033181\tablenotemark{\star} & 24.3 & 11 & 0.05(2) \nl
47~Tuc~R & 3.480463\tablenotemark{\star} & 24.4 & 15 & \nodata \nl
47~Tuc~S & 2.830\tablenotemark{\star}    & 24.4 & 9  & \nodata \nl
47~Tuc~T & 7.589\tablenotemark{\star}    & 24.6 & 17 & \nodata \nl
47~Tuc~U & 4.342827\tablenotemark{\star} & 24.3 & 8  & 0.06(1) \nl
47~Tuc~V & 4.810\tablenotemark{\star}    & 24.1 & 7  & \nodata \nl
47~Tuc~W & 2.352344\tablenotemark{\star} & 24.3 & 17 & \nodata \nl
\enddata
 
\tablecomments{Uncertainties in barycentric period and DM are one or
less in the last digits quoted.}

\tablenotetext{\star}{Binary pulsar (see Table~\ref{tab:bin}).}

\end{deluxetable}

As noted by a number of authors (\cite{mp86}; \cite{jk91}), standard
pulsar search analyses, which look for significant harmonics in the
power spectrum of a de-dispersed time series, suffer reductions in
sensitivity to pulsars in short-period binary systems.  The effect of
the binary motion is to cause a change in the apparent pulse frequency
during the integration, spreading the emitted signal power over a
number of spectral bins, thereby reducing the apparent signal-to-noise
ratio.  In order to recover the signal, it is necessary to transform
the time series to the rest frame of an inertial observer with respect
to the pulsar before carrying out the periodicity search.  This
transformation is readily achieved by applying the Doppler formula to
relate a time interval in the pulsar frame, $\tau$, to the
corresponding interval in the observed frame, $t$:
\begin{equation}
\label{eq:resamp}
	\tau(t) = \tau_0 ( 1 + v(t)/c ),
\end{equation}
where $v(t)$ is the observed radial velocity of the pulsar along the
line-of-sight, $c$ is the speed of light, and we have neglected terms
in $(v/c)$ of order higher than the first.  Ideally, if the orbital
parameters of the binary system are known, $v(t)$ can be calculated
from Kepler's laws.  For the purposes of a blind search, where the
orbital parameters are a priori unknown, assuming a Keplerian model for
$v(t)$ would require a five-dimensional search of all the parameter
space --- a computationally non-trivial task! The search can, however,
be considerably simplified by assuming that the orbital acceleration
$a$ is constant during the observation, i.e., $v(t)=at$.  This
assumption is commonly used in searches of this kind (\cite{and92}) and
turns out to be reasonable over a wide range of orbital parameters,
phases, and observation lengths.  We investigate this in more detail in
\S\ref{sec:sensb}.

Given a prescription for $v(t)$, the re-sampling process is
straightforward.  The time intervals in the new frame are calculated
from equation \ref{eq:resamp}.  New sample values are then created
based on a linear interpolation running over the original time series
(see also \cite{mk84}).  We choose to define the value of $\tau_0$ such
that $\tau$ is equal to the original sampling interval, $t_{\rm samp}$,
at the mid-point of the integration.  For the condition $v(t)=at$, it
follows that
\begin{equation}
	\tau_0 = \frac{t_{\rm samp}}{1+aT/2c},
\end{equation}
where $T$ is the integration time, 17.5\,min in our case.  This
condition guarantees that, under the approximations made here, the
number of samples in the corrected time series is the same as in the
original one.  The re-sampling process is relatively trivial in terms
of computation time and can be readily carried out for a number of
trial accelerations for each de-dispersed time series in order to
effect a search in ``acceleration space''.

In all the analyses carried out in this paper, each de-dispersed time
series was re-sampled and analyzed for a number of trial accelerations
in the range $|a|<30$\,m\,s$^{-2}$.  Most pulsars in presently known
binary systems typically undergo accelerations of up to about
5\,m\,s$^{-2}$.  There are however other binary systems where more
extreme accelerations have motivated us to search a larger range of
acceleration space.  These include the double neutron-star binary
PSR~B2127+11C in M15, originally discovered at a trial acceleration of
$-9.5$\,m\,s$^{-2}$ (\cite{agk+90}), and the eclipsing binary
PSR~B1744$-$24A in Terzan~5 (\cite{lmd+90}), where the maximum
line-of-sight acceleration is 33\,m\,s$^{-2}$.

The step size between each trial acceleration was 0.3\,m\,s$^{-2}$.
Some care is required when choosing this interval in order to strike a
compromise between unnecessary processing time incurred by oversampling
the parameter space and the loss of sensitivity caused by
undersampling.  Assuming that the constant acceleration approximation
adequately describes the Doppler shifts during the integration, it
follows that the number of spectral bins across which a signal will
drift due to the assumed trial acceleration being different from the
true value by an amount $\Delta a$ is given by $\Delta a \times
T^2/Pc$, where $P$ is the pulse period.  Our choice of step size thus
guarantees that any pulsar period or harmonic with a period greater
than 2\,ms will not drift by more than one spectral bin during the
17.5\,min integration.

Following the de-dispersing and re-sampling procedures outlined above,
each time series is analyzed independently according to standard pulsar
search techniques.  The analysis procedure is very similar to that
described in detail by Manchester et~al.~(1996).  \nocite{mld+96} In
brief, we compute discrete Fourier transforms of each time series, and
sum harmonically related spectral components for 1, 2, 4, 8, and 16
harmonics in turn, in order to maintain sensitivity to a variety of
pulse profile relative widths, or duty-cycles.  For each harmonic sum,
candidates are ranked in order of spectral signal-to-noise
ratios\footnote{Spectral signal-to-noise ratios are defined throughout
this paper to be the height of a feature in the amplitude spectrum
divided by the spectral rms.  In general, spectral signal-to-noise
ratios and ratios calculated from reconstructed time-domain pulse
profiles are different, although in practice, for relatively low values
--- relevant for search sensitivity calculations (\S\ref{sec:sens};
\S\ref{sec:sensb}) --- they are identical, and we use the nomenclature
$\sigma$ in both instances.  To estimate average flux densities
(\S\ref{sec:flux}), where some signal-to-noise ratios are large, we
carry out all computations in the time domain.}.  At this stage, all
spectral features with $\sigma \geq 6$ are stored, along with the
period of the signal.

Once computations are completed for the time series corresponding to
all trial accelerations for a 17.5\,min block of data, we plot $\sigma$
versus trial acceleration for multiple occurrences of each period,
within a suitable tolerance, detected above a pre-defined
``significance'' threshold.  This threshold can be estimated
theoretically from the statistics of the phase-space searched, but
practical considerations (such as the presence of radio-frequency
interference) result in our determining this value empirically.  We
generate plots such as that in Figure~\ref{fig:snr} for all cases where
at least one detection of a multiply detected period has $\sigma$ $\ge
9$.

\medskip
\epsfxsize=8truecm
\epsfbox{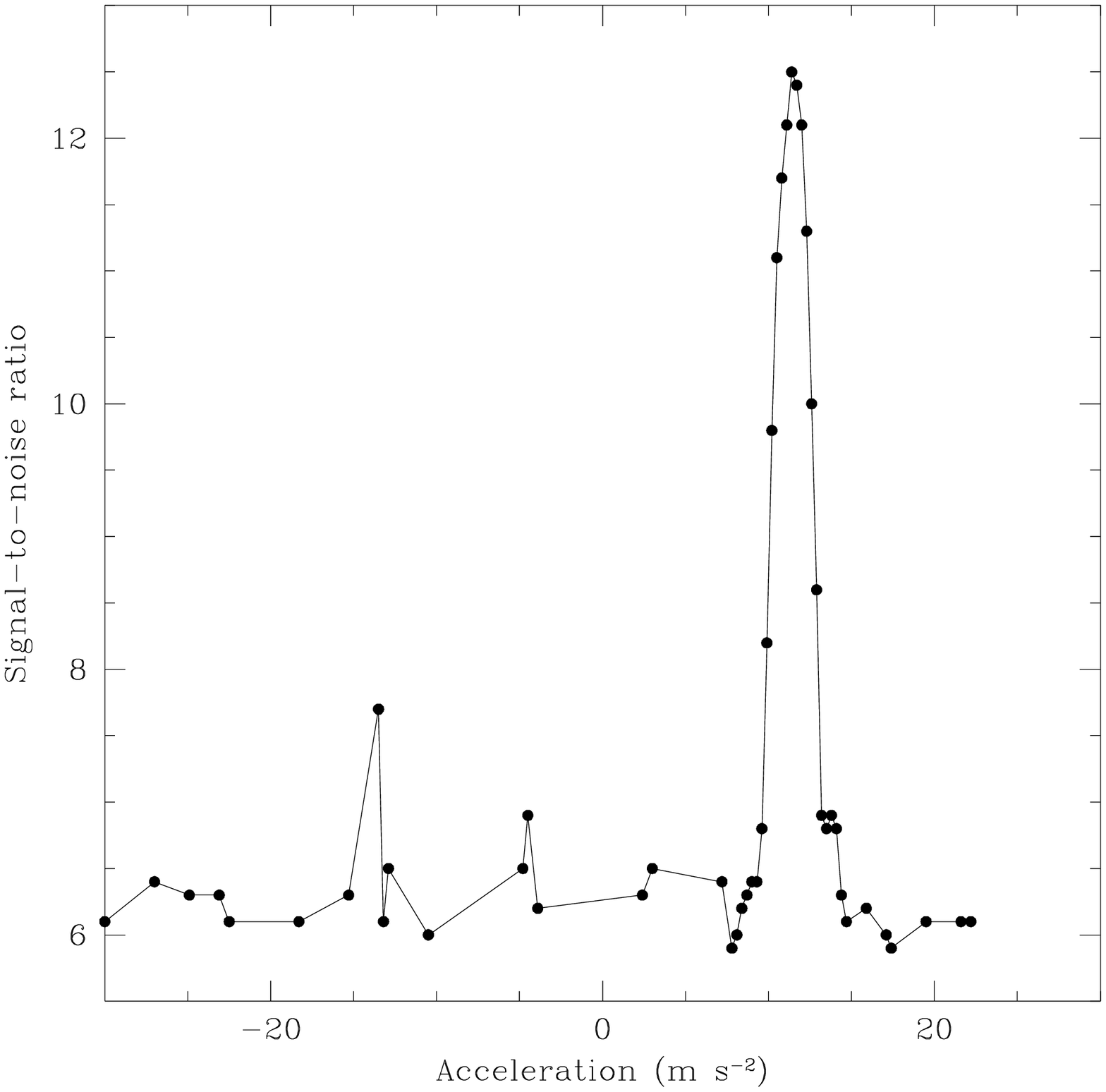}
\figcaption[snr.eps]{\label{fig:snr} Signal-to-noise ratio ($\sigma$)
versus trial acceleration for the discovery of 47~Tuc~R.  This plot
corresponds to one of seven 17.5\,min integrations, out of a total of
eight on MJD~50742, in which $\sigma$ in at least one trial
acceleration was higher than the threshold of 9.  Most points with
$\sigma \la $ 8 correspond to random noise rather than detection of the
pulsar.}
\bigskip

\section{Sensitivity}\label{sec:sens}

To estimate the sensitivity of our search we calculate the minimum flux
density $S_{\rm min}$ that a pulsar must have in order to be
detectable.  Following \nocite{dss+84} Dewey et~al.~(1984), this is
given by
\begin{equation}
\label{eq:smin}
   S_{\rm min} = \frac{\sigma \eta \, S_{\rm sys}}{\sqrt{n \Delta
        \nu T}} \left(\frac{w}{P-w} \right)^{1/2}.
\end{equation}
Here $\sigma$ is the signal-to-noise ratio threshold of the survey (9
in our case); $\eta$ is a constant that takes account of losses in
sensitivity due to digitization of the incoming data and other
inefficiencies in the system ($\sim$1.5 for our system); $S_{\rm sys}$
is the system equivalent flux density ($\sim$35\,Jy for the present
system in the direction of 47~Tuc); $n$ is the number of polarizations
summed (two in our case); $\Delta \nu$ is the total observing bandwidth
(288\,MHz); $T$ is the integration time (1049\,s); and $w$ and $P$ are
the pulse width and period respectively.

The observed pulse width $w$ in the above expression is in general
greater than the intrinsic width $w_{\rm int}$ emitted at the pulsar
because of the dispersion and scattering of pulses by free electrons in
the interstellar medium, and the post-detection integration performed
in hardware.  Interstellar scattering is negligible for most
observations toward 47~Tuc, which is relatively nearby and located well
away from the Galactic plane.  The observed pulse width for
unaccelerated pulsars will therefore be the convolution of the
intrinsic pulse width and broadening functions due to dispersion and
integration, and can be estimated from the quadrature sum
\begin{equation}
\label{eq:weff}
  w^2 = w_{\rm int}^2 + t_{\rm DM}^2 + t_{\Delta \rm DM}^2 + t_{\rm
          samp}^2.
\end{equation}
Here $t_{\rm DM}$ is the dispersion broadening across one filter bank
channel, $t_{\Delta \rm DM}$ is the overall pulse broadening due to the
difference between the single DM at which we de-disperse all data and
the true DM of a pulsar ($\Delta \rm DM$), and $t_{\rm samp}$ is the data sampling
interval (125\,$\mu$s in our case).  The dispersion broadening across a
single channel is calculated using the standard expression
\begin{equation}
\label{eq:dm}
 t_{\rm DM} = 8.3 \left(\frac{\rm DM}{{\rm cm}^{-3}\,{\rm pc}}\right)
                  \times \left(\frac{\Delta \nu_{\rm chan}}{\rm MHz}\right)
                  \times \left(\frac{\nu}{\rm GHz}\right)^{-3}\,{\rm \mu s},
\end{equation}
which is valid for $\Delta \nu_{\rm chan} \ll \nu$ as in our case where
$\Delta \nu_{\rm chan}=3$\,MHz and $\nu =1374$\,MHz.  The effective
time resolution for these 47~Tuc observations, calculated from
equation~\ref{eq:dm} and the sample time, at a DM of 25\,cm$^{-3}$\,pc,
is approximately 0.3\,ms at the center frequency of observations,
although it falls to 0.4\,ms at the low-frequency end.  However in many
cases the effective time resolution is limited by the $t_{\Delta \rm
DM}$ term in equation~\ref{eq:weff}, which amounts to
$0.9\,(\frac{\Delta{\rm DM}}{{\rm cm}^{-3}\,{\rm pc}})$\,ms at
1374\,MHz.

Figure~\ref{fig:sens} shows the search sensitivity as a function of
period for a 17.5\,min integration based on equations
\ref{eq:smin}--\ref{eq:dm}.  Here we have assumed a box-car pulse shape
and an intrinsic duty-cycle of 15\%, typical of many among the pulsars
known in 47~Tuc.  It can be seen that the present system is sensitive
to most pulsars in 47~Tuc with $P>2$\,ms and with average 20\,cm flux
densities above 0.3\,mJy during a 17.5\,min integration.

\medskip
\epsfxsize=8truecm
\epsfbox{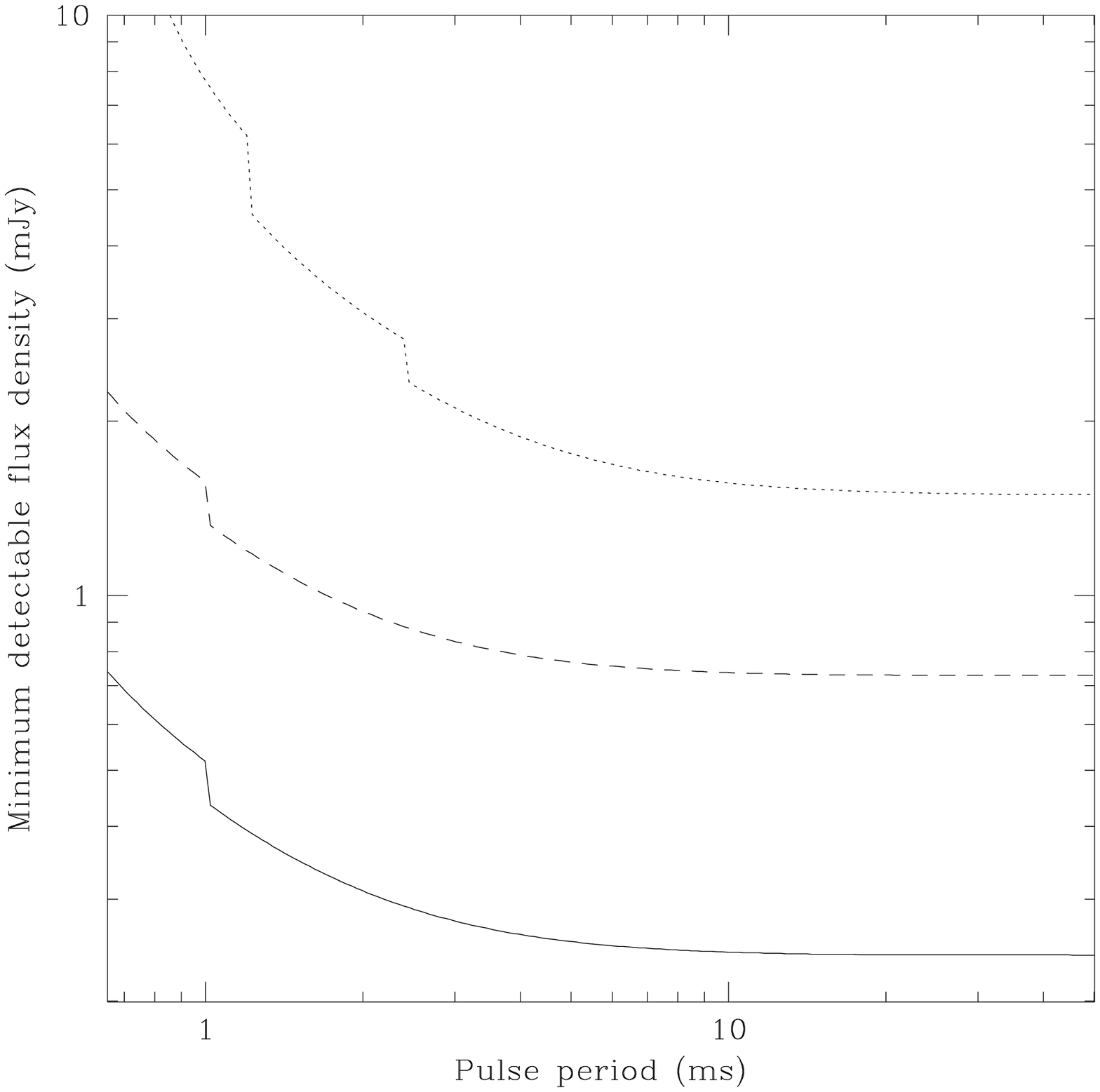}
\figcaption[sens.eps]{\label{fig:sens} Sensitivity as a function of
pulse period.  The solid curve is the sensitivity calculated for our
20\,cm observations, for an assumed pulsar duty-cycle of 15\%, analyzed
in individual 17.5\,min blocks.  The steps in the curves going from
right to left reflect loss of power in the frequency domain as
higher-order harmonics drop out of the spectrum.  The dotted curve is
an estimate of the sensitivity at 50\,cm for the observations reported
by Manchester et~al.~(1990, 1991), for an integration time of 42\,min.
The dashed curve scales the 20\,cm curve to a wavelength of 50\,cm
assuming a pulsar spectral index of $-1.6$.}
\bigskip

The improvement in sensitivity provided by the new 20\,cm system is
apparent when comparing its sensitivity curve to that based on a
42\,min observation assuming the parameters of the 50\,cm system used
by Manchester et~al.~(1990, 1991) \nocite{mld+90,mlr+91} to discover
pulsars 47~Tuc~C--M, also shown in Figure~\ref{fig:sens}.  For flat
spectrum sources one may compare the 20 and 50\,cm curves directly.  A
more realistic comparison between the two searches is shown by the
scaled version of the present 20\,cm curve assuming the typical pulsar
spectral index of $-1.6$.  With this assumption we see that the present
system represents as much as a threefold improvement in sensitivity for
a $\sim 2$\,ms pulsar such as 47~Tuc~J.

The sensitivity calculations described in this section take no account
of the effects of the degradation in $\sigma$ of any binary pulsar
whose apparent period may change significantly during the 17.5\,min
integration.  We investigate this issue in detail in
\S\ref{sec:sensb}.

\section{Results}\label{sec:res}

We have reduced all data listed in Table~\ref{tab:obs} in the manner
described in \S\ref{sec:data}.  In this table we list the number of
17.5\,min integrations analyzed on each day, giving both the number of
daily integrations in which the search code detected the pulsar ---
i.e., in which $\sigma_{\rm search}$ $\ge9$ --- and the average daily
$\sigma$, scaled to a 17.5\,min integration, obtained from summing all
data for that day using the pulsar's ephemeris.  We detected some 50
distinct periods for which the $\sigma$ vs. trial acceleration plots
are broadly similar to that of Figure~\ref{fig:snr}, i.e., have a
clearly defined and significant maximum in trial acceleration.  Eleven
of these we identified with previously known pulsars in 47~Tuc.

\subsection{Previously Known Pulsars}\label{sec:resold}

The detection rate for the previously known pulsars ranges from 10\%
(47~Tuc~N), to 90\% (47~Tuc~C and J), with all but two of the 11
pulsars detected on at least 25\% of the observing days.  In contrast,
at 50\,cm, all but 47~Tuc~C are detected less than 25\% of the time,
and the same is true at 70\,cm for seven of the pulsars
(\cite{rob94}).  The detection rates with the new observing system are
clearly much higher than with past observations, leading to improved
parameters for most pulsars.

In Table~\ref{tab:3sol} we present ``phase-connected'' timing solutions
for three previously known pulsars, obtained from 20\,cm data collected
since 1997 October.  To determine these we first make time-of-arrival
(TOA) estimates by cross-correlating pulse profiles, obtained by
summing several minutes' worth of data for a pulsar on a given day,
with a high-$\sigma$ ``standard profile'', and adding the time delay to
the start-time of the integration, appropriately translated to the
mid-point of the integration.  We then use the TOAs, and a model
describing the pulsars, including astrometric, spin, and (where
relevant) binary parameters, to obtain improved parameters.  These are
derived by minimizing the timing residuals (difference between observed
and computed pulse phases) in a least-squares sense.  This procedure is
implemented in the {\sc tempo} software package 
({\tt http://pulsar.princeton.edu/tempo}).

\begin{deluxetable}{llll}
\scriptsize
\tablecaption{\label{tab:3sol}Timing solutions of three pulsars.}
\tablecolumns{4}
\tablehead{
\colhead{Parameter}     & 
\colhead{J0023$-$7204C} & 
\colhead{J0024$-$7204D} & 
\colhead{J0023$-$7203J} \nl}
\startdata
Right ascension, $\alpha$ (J2000)  \dotfill   & 
$00^{\rm h}23^{\rm m}50\fs350(1)$  & $00^{\rm h}24^{\rm m}13\fs877(2)$ &
$00^{\rm h}23^{\rm m}59\fs4057(6)$ \nl
Declination, $\delta$ (J2000)      \dotfill   &
$-72\arcdeg04'31\farcs486(3)$      & $-72\arcdeg04'43\farcs845(9)$  &
$-72\arcdeg03'58\farcs781(2)$      \nl
Period, $P$ (ms)                   \dotfill   &
5.7567800118(2)   & 5.3575732858(1)   & 2.100633545861(1) \nl
Period derivative, $\dot P$        \dotfill   & $-5.1(1)\times10^{-20}$ &
$-0.31(4)\times10^{-20}$ & $-0.96(1)\times10^{-20}$ \nl
Epoch of period (MJD)              \dotfill   &
47858.5 & 48040.7 & 51000.0 \nl
Dispersion measure, DM (cm$^{-3}$\,pc) \dots  &
24.6    & 24.7    & 24.6    \nl
Orbital period, $P_b$ (d)          \dotfill   &
\nodata & \nodata & 0.120664939(1)  \nl
Projected semi-major axis, $x$ (s) \dotfill   &
\nodata & \nodata & 0.040411(1)     \nl
Eccentricity, $e$                  \dotfill   &
\nodata & \nodata & $<0.0001$ \nl
Longitude of periastron, $\omega$  \dotfill   &
\nodata & \nodata & 0.0 \nl
Time of periastron, $T_0$ (MJD)    \dotfill   &
\nodata & \nodata & 51000.765678(1) \nl
\tableline
Post-fit rms timing residual ($\mu$s) \dotfill & 10 & 19 & 9 \nl
$\alpha - \alpha_{\rm 47~Tuc}$ \dotfill &
$-15\fs6$ & $8\fs0$ & $-6\fs5$ \nl
$\delta - \delta_{\rm 47~Tuc}$ \dotfill &
$19\farcs6$ & $7\farcs3$ & $52\farcs3$ \nl
Pulsar--cluster center offset (core radii) \dotfill & 6 & 3 & 5 \nl
\enddata

\tablecomments{Figures in parentheses represent uncertainties in the
last digits quoted, given for all fitted parameters.}

\end{deluxetable}

The timing solutions for 47~Tuc~C and D match those first presented by
Robinson et~al.~(1995)\nocite{rlm+95}, obtained from 50 and 70\,cm
data.  The phase-coherent solution for 47~Tuc~J is a new one.  The
small limit on its orbital eccentricity is perhaps somewhat surprising,
since it suggests a lack of strong recent interactions with other
stars, while its position well outside the cluster core tends to
suggest the opposite.  However it is also conceivable that tidal
dissipation in a Roche lobe-filling companion could circularize the
orbit quickly (cf. \S\ref{sec:ecl}).  We remark in passing that a
20\,cm continuum map of the central regions of 47~Tuc shows three
relatively bright, variable, point sources (\cite{fg99}).  Two of these
are clearly associated with 47~Tuc~C and D.  The third, whose position
differs from that listed for 47~Tuc~J in Table~\ref{tab:3sol} by
$0\fs07(3)$ in right ascension and $0\farcs3(2)$ in declination, and
whose average flux density of 0.6\,mJy matches our average of 0.5\,mJy
(\S\ref{sec:flux}), can now be identified with 47~Tuc~J.

There is no evidence that 47~Tuc~L is a binary pulsar (a possibility
according to Robinson et~al.~1995\nocite{rlm+95}).  We have detected
this pulsar on 11 separate occasions spanning over 500 days
(Table~\ref{tab:obs}), and the barycentric period is constant (see
Table~\ref{tab:spin}, which lists the pulse periods, DMs, and observed
duty-cycles of all pulsars detected in our observations).

Finally, we were also able to determine the eccentricity for 47~Tuc~E,
and obtain a binary solution for 47~Tuc~H.  The very significant
eccentricities for these systems (Table~\ref{tab:bin}) are unique among
this class of binaries: in the disk of the Galaxy such systems have
$e\la10^{-5}$ (\cite{cam98}).  The eccentricities are presumably fossil
evidence of interactions near the dense cluster core, and may be due to
perturbations by passing stars, or stellar exchange interactions (see
\cite{rh95} and \cite{hr96} for details).  It will be interesting to
see if other wide systems in this cluster have large eccentricities.

\vspace{-2mm}
\subsection{Newly Discovered Pulsars}\label{sec:resnew}

Nine of the remaining periods detected in our data appeared in at least
two 17.5\,min integrations on one day, with some being detected in most
integrations.  For each detection of such a period, we de-dispersed the
raw data at several trial values of DM about the nominal DM.  In all
nine cases there was a clearly defined maximum signal-to-noise ratio as
a function of DM, ensuring that the periodic signals were truly
dispersed.  After viewing the corresponding time-domain pulse profiles,
and judging them suitably ``pulsar-like'', we felt secure in
identifying each of these nine periods with previously unknown pulsars
in 47~Tuc.  Table~\ref{tab:rare} lists detections for the newly
discovered pulsars that are seen only occasionally.

For all nine of the newly detected pulsars, the period changed within
an observing day in a manner consistent with the Doppler shifts arising
in binary systems.  Only three of the nine pulsars (47~Tuc~O, Q and U)
were detected on many occasions (Table~\ref{tab:obs}).  Fortunately,
three of the remaining six (47~Tuc~P, R, and W) have such short orbital
periods that it was possible to cover more than an entire orbit on
their discovery days, and hence establish good orbital parameters (see
Fig.~\ref{fig:vel}).  Binary parameters for all pulsars are summarized
in Table~\ref{tab:bin}.

Most of the newly discovered pulsars appear to be similar in their
binary characteristics to the previously known pulsars in the cluster.
For instance, 47~Tuc~O, P, and R have short orbital periods ($P_b\sim
1.5$--5.5\,h) and low companion masses ($m_2\sim0.03$\,M$_\odot$),
similar to the previously known 47~Tuc~I and J.  Although the period of
47~Tuc~R, at 96\,min, is considerably shorter than that of 47~Tuc~J,
there is no reason to suspect that these systems are fundamentally
different.

In contrast, 47~Tuc~Q and U are similar to 47~Tuc~E and H, with
$m_2\sim0.2$\,M$_\odot$; the companion stars are presumably low-mass
helium white dwarfs much like those found in similar systems in the
Galactic disk (e.g., \cite{lfc96}).  The binary periods for 47~Tuc~Q
and U are somewhat shorter than the corresponding ones for 47~Tuc~E and
H (0.4--1.2\,d vs. 2.3\,d; see Table~\ref{tab:bin}), but we believe
this betrays no fundamental difference between the various systems.  We
have not yet obtained orbital parameters for 47~Tuc~S and T, but from
the observed projected velocity change during the discovery days ($-5$
and 12\,km\,s$^{-1}$ over 2\,h, respectively; see Fig.~\ref{fig:vel})
we believe they also fall into this higher-mass group.

It thus appears the vast majority of the binary pulsars known to date
in 47~Tuc fall into one of two classes.  However, the two pulsars
discovered most recently, 47~Tuc~V and W, are not so easily
categorized.

\medskip
\epsfxsize=8truecm
\epsfbox{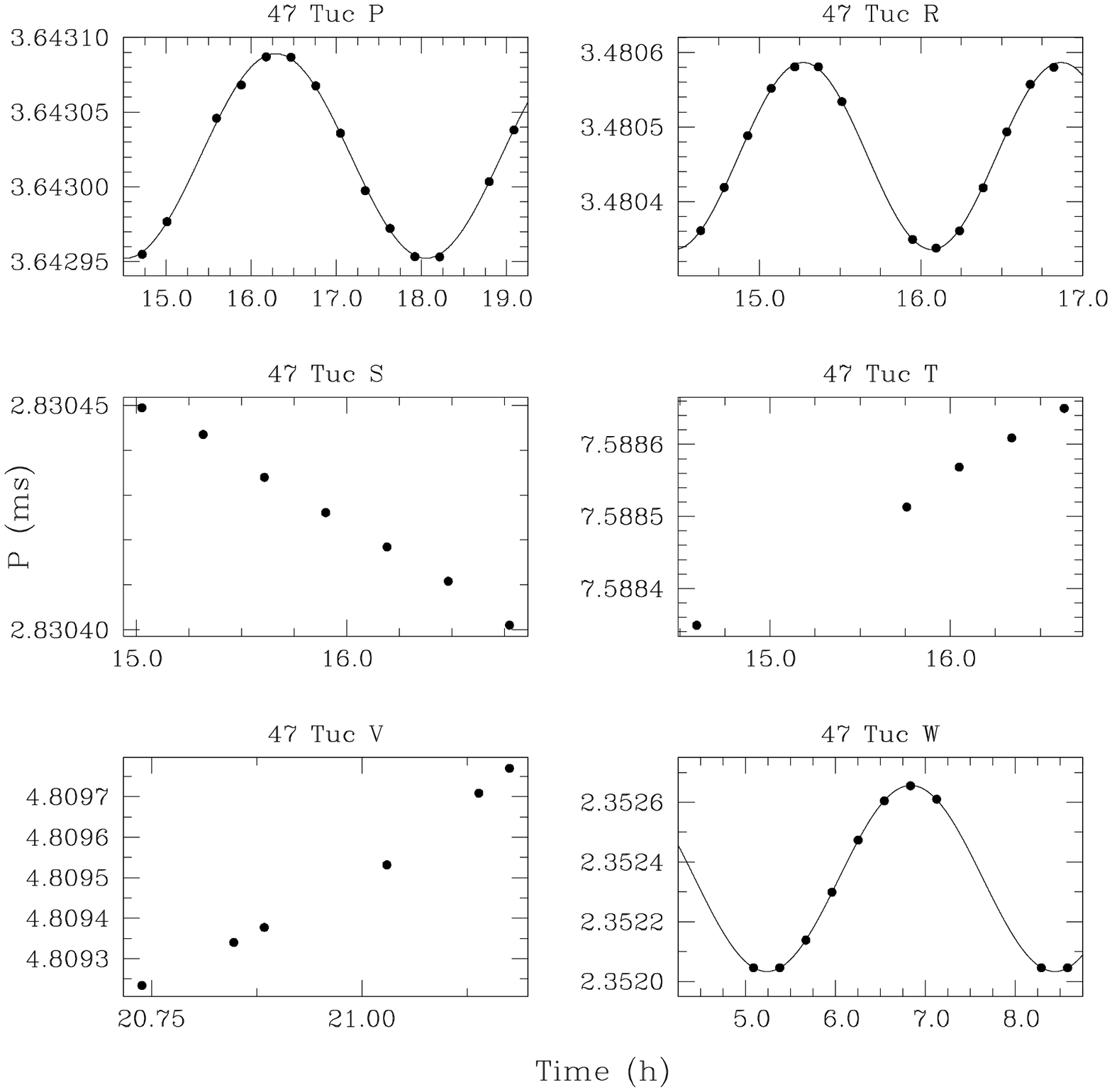}
\figcaption{\label{fig:vel} Observed barycentric periods for six newly
discovered pulsars that are seldom detected.  Each period is based on
17.5\,min of data, with the exception of 47~Tuc~R (8.7\,min) and
47~Tuc~V (2.2\,min), and corresponds to $\sigma\ge6$.  Times are given
in UT on discovery day (see Table~\protect\ref{tab:rare}).}
\bigskip

\begin{deluxetable}{lllllll}
\scriptsize
\tablecaption{\label{tab:bin}Binary parameters.}
\tablecolumns{7}
\tablehead{
\colhead{Pulsar}   & 
\colhead{$P_b$}    & 
\colhead{$x$}      & 
\colhead{$T_0$}    &
\colhead{$\omega$} &
\colhead{$e$}      &
\colhead{$m_2\tablenotemark{\star}$}  \nl
\colhead{}         &
\colhead{(d)}      &
\colhead{(s)}      &
\colhead{(MJD)}    &
\colhead{($\deg$)} &
\colhead{}         &
\colhead{(${\rm M_\odot}$)} \nl}
\startdata
\multicolumn{6}{c@{\vspace{1mm}}}{Previously Known} \nl
\tableline
47~Tuc~E & 2.2568448 & 1.98184  & 51004.05      & 219    & 0.0003  & 0.19 \nl
47~Tuc~H & 2.3576966 & 2.15281  & 51000.9735    & 110    & 0.071   & 0.20 \nl
47~Tuc~I & 0.2297923 & 0.03846  & 50740.5794    &   0.0  & 0.0     & 0.02 \nl
47~Tuc~J & 0.1206649 & 0.04041  & 51000.7657    &   0.0  & 0.0     & 0.03 \nl
\tableline
\multicolumn{5}{c@{\vspace{-1mm}}}{} \nl
\multicolumn{6}{c@{\vspace{1mm}}}{Newly Discovered} \nl
\tableline
47~Tuc~O & 0.1359743 & 0.04515  & 51000.0212    &   0.0  & 0.0     & 0.03 \nl
47~Tuc~P & 0.1472    & 0.0380   & 50689.6789    &   0.0  & 0.0     & 0.02 \nl
47~Tuc~Q & 1.1890840 & 1.46219  & 51002.175     &   0.0  & 0.0     & 0.21 \nl
47~Tuc~R & 0.0662    & 0.0334   & 50742.6365    &   0.0  & 0.0     & 0.03 \nl
47~Tuc~S &\tablenotemark{\dagger}&\nodata &\nodata &\nodata &\nodata
&\nodata\nl
47~Tuc~T &\tablenotemark{\dagger}&\nodata &\nodata &\nodata &\nodata
&\nodata\nl
47~Tuc~U & 0.4291057 & 0.52696  & 51002.645     &   0.0  & 0.0     & 0.15 \nl
47~Tuc~V & $\sim0.2?$ &\nodata &\nodata &\nodata &\nodata &\nodata\nl
47~Tuc~W & 0.1330    & 0.2435   & 51214.950     &   0.0  & 0.0     & 0.15 \nl
\enddata

\tablecomments{Uncertainties in all fitted parameters are one or less
in the last digits quoted.}

\tablenotetext{\star}{Assumes a pulsar mass of 1.4\,${\rm M_\odot}$ and
inclination angle for the orbit of $i=60\arcdeg$.}

\tablenotetext{\dagger}{About one to several days.}

\end{deluxetable}

Pulsar 47~Tuc~V has been detected on two different days, at seemingly
identical orbital phases (with period increasing as a function of time,
hence on the ``near-side'' of its companion --- see
Fig.~\ref{fig:vel}), and it displays an apparent orbital velocity
change of up to $\sim 80$\,km\,s$^{-1}$ in under one hour.  The
observed change in period during this time leads us to believe that the
orbital period is several hours, suggesting a relatively massive
companion.  Another unusual aspect of this system is that it was
detected on both occasions with relatively large signal-to-noise ratios
in several 2\,min sub-integrations, but interspersed with
non-detections (Fig.~\ref{fig:vel}), as if it were being irregularly
eclipsed, despite its orbital phase\footnote{In practice it is usually
easy to distinguish among interstellar scintillation and eclipses by a
companion as the cause for the temporary non-detection of a pulsar.
Eclipses tend to occur with the pulsar roughly centered at superior
conjunction, while scintillation has no preferred orbital phase.  Also,
scintillation has characteristic time-scales and amplitudes.  No other
pulsar in 47~Tuc displays the deep intensity modulations on time-scales
of about a minute observed for 47~Tuc~V, and we therefore suggest the
eclipse interpretation.}.

Pulsar 47~Tuc~W, with an orbital period of 3.2\,h, may resemble
short-period binary systems such as 47~Tuc~J, and the pulsed radio
signals are apparently eclipsed for a fraction of its orbit, as in some
of these systems (see \S\ref{sec:ecl}).  However its implied companion
mass, at $\sim0.15$\,M$_\odot$, is a factor of $\sim 6$ higher than
that of those systems, more typical of a low-mass white-dwarf.  But if
the companion were indeed a white dwarf, we would not expect eclipses.
We consider it possible that this system is somewhat akin to
PSR~B1744$-$24A, located in the cluster Terzan~5, with $P_b\sim
1.8$\,h, $m_2\sim0.10$\,M$_\odot$, and displaying irregular eclipses
(\cite{lmd+90}; \cite{nt92}).

There remain about 30 periodic signals that displayed several
pulsar-like characteristics, but that were detected on only one
integration of one day.  We regard several of these as good pulsar
candidates, potentially to be confirmed by future observations.

\begin{deluxetable}{llll}
\scriptsize
\tablecaption{\label{tab:rare}Rarely detected pulsars.}
\tablecolumns{4}
\tablehead{
\colhead{Pulsar}              & 
\colhead{Date}                & 
\colhead{Number of}           & 
\colhead{Detections/$\sigma$} \nl
\colhead{}                    & 
\colhead{(MJD)}               & 
\colhead{integrations}        &
\colhead{}                    \nl}
\startdata
47~Tuc~P & 50689 &   16 & 5/6.3 \nl
47~Tuc~R & 50742 &    8 & 7/8.9 \nl
47~Tuc~S & 50741 &    8 & 3/9.9 \nl
47~Tuc~T & 50746 &    8 & 4/9.3 \nl
         & 51005 &    8 & 0/5.4 \nl
         & 51040 &    4 & 4/25.8\nl
47~Tuc~V & 51012 &    8 & 2/3.4 \nl
         & 51055 &   16 & 3/4.4 \nl
47~Tuc~W & 51214 &   16 & 9/10.0\nl
\enddata

\end{deluxetable}

\section{Discussion}\label{sec:disc}

\subsection{Survey Limitations}\label{sec:limit}

Although the recent observations of 47~Tuc have been successful beyond
our most optimistic expectations, it is important to keep in mind the
many and severe selection effects inherent in the data and analyses
reported on in this paper.  The greatest of these, particularly for
short-period binary systems, has to do with the extent of parameter
space that has been covered by our ``acceleration searches''.

\subsubsection{Sensitivity to Binary Pulsars}\label{sec:sensb}

In addition to the improvements in sensitivity over previous searches
of 47~Tuc enabled by the new 20\,cm observing system, use of an
acceleration search has significantly improved the ability to find
binary pulsars.  Prior to this survey only four out of 11 (36\%)
millisecond pulsars known in 47~Tuc were members of binary systems.
The new discoveries bring the number of binary pulsars in 47~Tuc to 13
(65\% of the total).  We consider it highly significant that all of the
new millisecond pulsars reported here are members of binary systems.
These discoveries are in spite of the fact that, as demonstrated in
\S\ref{sec:sens}, the sensitivity to isolated pulsars is significantly
improved over the previous surveys by Manchester et~al.~(1990, 1991)
\nocite{mld+90,mlr+91} and Robinson et~al.~(1995)\nocite{rlm+95}.  We
note that three of the five brightest pulsars in 47~Tuc at 20\,cm are
isolated (Table~\ref{tab:spin}); if significant, this may suggest
that isolated pulsars in 47~Tuc have a relatively high intrinsic
luminosity cutoff.  Also, isolated pulsars in 47~Tuc might have, on
average, a larger spectral index than binary pulsars.  Finally, the
specific incidence of isolated pulsars in 47~Tuc might be low.  A
combination of these factors could explain why our more sensitive
survey has failed to find additional isolated pulsars.

The improved sensitivity of the acceleration search to short-period
binaries is demonstrated by the fact that four of the nine binary
systems we found, 47~Tuc~P, R, S and V, would not have been discovered
without the use of the acceleration code since, in each case, the
signal-to-noise ratio never rose above the threshold of 9 at a trial
acceleration of zero\footnote{In effect this suggests that without the
acceleration code we would have discovered five new pulsars, an
increase of $\sim 50\%$ over the previously known population.  This is
surprising, considering the factor of $\sim 3$ improvement in
sensitivity claimed in \S\ref{sec:sens}.  This estimate relied on
assuming a spectral index of $-1.6$ for pulsars in 47~Tuc.  Perhaps
this assumption is not valid, and the average spectral index for
millisecond pulsars in 47~Tuc is smaller than $-1.6$.}.  Despite
searching an acceleration range of $|a|<30$\,m\,s$^{-2}$, most of the
new discoveries and detections of previously known pulsars occurred at
$|a|<5$\,m\,s$^{-2}$.  Three notable exceptions are 47~Tuc~R, V, and W,
that displayed maximum line-of-sight accelerations of 11.4, 23.7, and
21.3\,m\,s$^{-2}$ respectively (e.g., Fig.~\ref{fig:snr}).

While it is clear that this survey has improved the knowledge of the
true fraction of binary pulsars in 47~Tuc, an outstanding question that
remains is the extent of binary parameter space probed by the present
search strategy.  In this section we attempt to answer this by
exploring the limits of the present search with respect to a variety of
short orbital period binary systems.  In what follows, when we use the
term ``acceleration search'' we are explicitly referring to the
one-dimensional (i.e., constant acceleration) approximation outlined in
\S\ref{sec:data}.

For the purposes of this discussion it is useful to note that most
observed binary pulsar systems in 47~Tuc fall broadly into two
categories: those with orbital periods of order 0.4--2.3\,d and
companion masses $\sim 0.2$ M$_{\odot}$ (47~Tuc~E, H, Q, U, and
probably S and T) and 47~Tuc~I, J, O, P, and R, which are characterized
by shorter orbital periods (1.5--5.5\,h) and lighter companions
($m_2\sim0.03$\,M$_{\odot}$).  (The nature of 47~Tuc~V and W is unclear
at present.) We shall refer to these types here respectively as the
``normal'' and ``short-period'' binary systems, investigating the
sensitivity of our search to both types in turn.  In addition, since
all the presently known binary systems have essentially circular
orbits, we also investigate the search sensitivity to putative binary
systems in which the pulsar is in an eccentric orbit about a more
massive companion.

The first, and really only, detailed study of the degradation in
sensitivity of a radio pulsar search for short orbital period systems
was carried out by \nocite{jk91} Johnston \& Kulkarni (1991; hereafter
JK).  By calculating the decrease in amplitude of the signal in the
Fourier power spectrum for a binary pulsar relative to a solitary one,
and mathematically describing the signal recovery technique involved in
the acceleration search, JK were able to quantify the degradation in
sensitivity of pulsar search codes with and without acceleration
schemes of the type used here.  As their main example, JK considered a
1000\,s observation of a 1.4\,M$_{\odot}$ pulsar in a circular orbit
with a 0.3\,M$_{\odot}$ companion for a wide variety of orbital and
pulse periods.  Given that our 17.5\,min analyses correspond to a
similar integration time, JK's results are highly relevant to this
discussion.  In addition, the 0.3\,M$_{\odot}$ companion considered by
JK closely resembles the class of normal binary systems defined above.

We focus on Figure~4 of JK, which shows the net gain in sensitivity of
an acceleration search when making multiple observations of the same
system.  In their example, JK assumed five independent observations of
the binary at random orbital phases.  From this figure we see that, for
a 10\,ms pulsar, the acceleration search allows one to detect binary
systems with orbital periods of order four times shorter than a
standard pulsar search code.  The gain increases still further for
shorter period pulsars, where the blurring of harmonics in the power
spectrum becomes even more pronounced.  For a 3.5\,ms pulsar, similar
to 47~Tuc~E and H, we infer from Figure~4 of JK that the acceleration
analysis allows the detection of a binary with an orbital period about
six times shorter than the standard search.

The results of JK are not directly applicable to the population of
short-period binaries in 47~Tuc.  Presently, the 3.48\,ms pulsar 47~Tuc
R, with an orbital period of 96\,min, is the shortest orbital period
system of this type.  In order to estimate the sensitivity of our
search code to even shorter period binaries with low-mass companions
like 47~Tuc~R, we generated a number of time series containing a
synthetic 3.48\,ms pulsar whose period was modulated as in a circular
Keplerian orbit with a low-mass companion.  We considered four separate
cases corresponding to orbital periods of 90, 60, 30 and 15\,min.  The
first value considered is similar to that of 47~Tuc~R; the last value
is close to that of the shortest orbital period known for any binary
system: the 11\,min orbit of the X-ray burst source X1820$-$303 in
the globular cluster NGC~6624 (\cite{spw87}).

In each case the mass function was held constant at $8.68\times10^{-6}$
M$_{\odot}$ (similar to that of 47~Tuc R).  To quantify the improvement
gained by using the acceleration code, we generated a control time
series containing a fake 3.48\,ms pulsar with a fixed period but
otherwise identical to the other simulations.  In a similar sense to
JK, we define the efficiency factor $\gamma$ as the detected $\sigma$
of the binary pulsar divided by that of the control pulsar.  In order
to compare the relative efficiency of the non-accelerated and
accelerated analyses, separate values of $\gamma$ were computed:
$\gamma_0$ is the efficiency for the non-accelerated analysis, and
$\gamma_A$ is the equivalent parameter for the acceleration search in
the analysis of the same data, performed as described in
\S\ref{sec:data}.  By definition $0 \leq \gamma \leq 1$, with a value
of unity signifying that no loss of signal occurred.  In each case, the
simulations were performed over all orbital phases to establish the
response of the search code to the full range of orbital
accelerations.

The results of the simulations are summarized in Table~\ref{tab:sim},
where we list the range of efficiencies, as well as the mean values
averaged over all orbital phases.  As expected, the efficiencies for
both the normal ($\gamma_0$) and acceleration search code ($\gamma_A$)
are strong functions of orbital phase.  In Table~\ref{tab:sim} we also
list the ``improvement factor'' $I\equiv\gamma_A/\gamma_0$, the net
gain in signal achieved by employing the acceleration search, averaged
by time spent at each orbital phase.  These simulations show a
significant improvement in sensitivity gained when using an
acceleration search.  In particular for orbital periods as short as one
hour (cases 1 and 2) the efficiency of the acceleration search is at
least 70\% for any orbital phase, more than twice the efficiency of
the normal search code.

\begin{deluxetable}{ccclcccccccc}
\scriptsize
\tablecaption{\label{tab:sim}Search code tests on simulated data.}
\tablecolumns{12}
\tablehead{
\colhead{Case}              &
\colhead{Period}            &
\colhead{$P_b$}             & 
\colhead{$f$}               & 
\colhead{$e$}               & 
\colhead{$\gamma_{0, {\rm min}}$} &
\colhead{$\overline{\gamma_0}$}   & 
\colhead{$\gamma_{0, {\rm max}}$} &
\colhead{$\gamma_{A, {\rm min}}$} &
\colhead{$\overline{\gamma_A}$}   & 
\colhead{$\gamma_{A, {\rm max}}$} &
\colhead{$I$}               \nl
\colhead{}                  &
\colhead{(ms)}              &
\colhead{(min)}             &
\colhead{(${\rm M_\odot}$)} &
\colhead{}                  &
\colhead{(\%)}              &
\colhead{(\%)}              &
\colhead{(\%)}              &
\colhead{(\%)}              &
\colhead{(\%)}              &
\colhead{(\%)}              &
\colhead{}                  \nl}
\startdata
1 & 3.48 & 90 & 8.68$\times10^{-6}$ & 0.0 & 33&48&75 & 80&88&99 & 1.8 \nl
2 & 3.48 & 60 & 8.68$\times10^{-6}$ & 0.0 & 26&39&58 & 68&75&86 & 1.9 \nl
3 & 3.48 & 30 & 8.68$\times10^{-6}$ & 0.0 & 24&33&41 & 38&45&51 & 1.4 \nl
4 & 3.48 & 15 & 8.68$\times10^{-6}$ & 0.0 & 18&23&28 & 30&32&35 & 1.4 \nl
5 & 34.8 & 60 & 0.13                & 0.7 & 0 &48&78 & 18&63&92 & 1.3 \nl
6 & 34.8 & 30 & 0.13                & 0.7 & 0 &33&63 & 20&43&69 & 1.3 \nl
\enddata

\tablecomments{For each case considered, the columns list the assumed
pulse period, orbital period, mass function, and orbital eccentricity.
The remaining columns show the range and average values of efficiences
of the search code for a normal analysis, $\gamma_0$, and using an
acceleration search, $\gamma_A$.  The averaged ``improvement factor''
$I\equiv\gamma_A/\gamma_0$ is also given.  Averaging over orbital
phases has been appropriately weighted by the time spent at each
phase.}

\end{deluxetable}

For the shortest orbital periods of this type considered (cases 3 and
4), the acceleration search continues to be more efficient than the
normal search although the improvement factor is smaller.  In these
cases the integration is becoming an increasingly significant fraction
of the orbital period.  As a result, the constant acceleration
approximation used in the search (\S\ref{sec:data}) ceases to hold.
Among possible improvements to this strategy, the most straightforward
approach is to perform one-dimensional acceleration searches on shorter
segments of data.  One can also include the acceleration derivative
term in the prescription for $v(t)$.  Although this necessarily adds a
dimension to the search, it may provide a dramatic improvement in
coverage of orbital parameter space for radio pulsars with binary
characteristics akin to X1820$-$303.

Finally, in Table~\ref{tab:sim} we also summarize the results of
simulations carried out to test the efficiency of our code at detecting
a putative relativistic binary double-neutron star system in a highly
eccentric short-period orbit (cases 5 and 6).  As would be expected,
for certain orbital phases a normal search has no hope of detecting
such systems, while the acceleration search maintains a detection
efficiency of at least 20\% at all orbital phases.  Our search would
therefore have found a relatively bright relativistic binary pulsar in
47~Tuc at any orbital phase, but none was detected.  This provides some
empirical evidence that the population of pulsars in the central
regions of 47~Tuc is composed exclusively of bona fide millisecond
pulsars: those with rotation periods under $\sim 10$\,ms, and in
relatively circular orbits with low-mass companions, or with no
companions at all.

\subsubsection{Other Selection Effects}\label{sec:selection}

An interesting issue from the point of view of the origin and evolution
of the pulsars in 47~Tuc is their intrinsic pulse period distribution.
The observed distribution shown in Figure~\ref{fig:phist} spans a
remarkably narrow range and suggests that there may be a deficit of
pulsars below 2\,ms.  This interpretation should be viewed with some
caution, however, given the small number statistics, and it is clear
from Figure~\ref{fig:sens} that we begin to lose sensitivity to pulsars
with $P\la2$\,ms.  Presently, therefore, the observations do not
unambiguously address the question of what the shortest pulsar periods
might be.  Pulsar 47~Tuc~W, for example, was originally detected at a
period of 1.18\,ms, showing that we are sensitive to some rather short
periods, but to be truly sensitive to $P\la1$\,ms, a different
experiment with higher time and frequency resolution will have to be
carried out.

\medskip
\epsfxsize=8truecm
\epsfbox{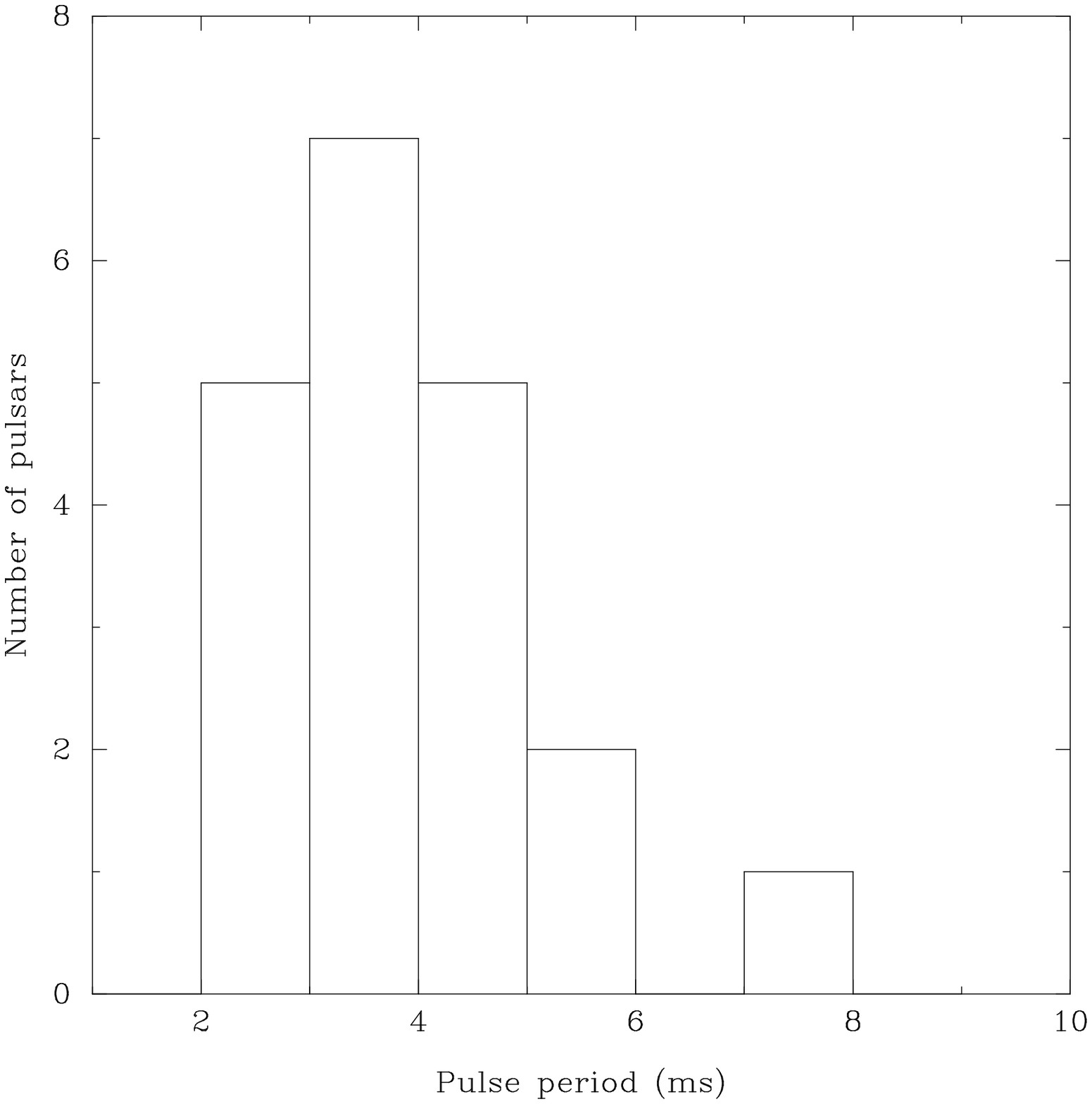}
\figcaption{\label{fig:phist} The observed period distribution of the 20
pulsars known in 47~Tuc.  It is presently unclear to what extent the
apparent deficit of pulsars with periods below 2\,ms is a result of
period-dependent selection effects (see
\protect\S\protect\ref{sec:selection}).}
\bigskip

It is likely that the vast majority of all pulsars in 47~Tuc are to be
found within the area covered by the $14'$-wide 20\,cm telescope beam
pattern.  This is in keeping with expectations based on the radial
density profile of 47~Tuc (\cite{dps+96}) from which we infer that
almost 70\% of the stars within the tidal radius of the cluster
(Table~\ref{tab:47tuc}) fall within the telescope beam area.  This is
additionally suggested by the observation that the pulsars for which a
timing solution has been obtained are all located within a region with
linear size $\sim 2'$, and that we have re-detected all pulsars
discovered at lower frequencies, even though the area subtended on the
sky by the 20\,cm beam pattern is at most one-sixth that within which
the first 11 pulsars were found.  Nevertheless it is possible that some
pulsars have been ejected from the central regions of the cluster, as
is thought to be the case for M15 (\cite{ps91}), and efforts to widen
the search region might prove instructive.

There are a number of additional selection effects having to do with
the manner in which we have reduced the data so far.

For isolated pulsars, the sensitivity limits might be improved by a
factor of up to $\sim \sqrt{16}$ by reducing the entire 4.66\,h data
sets in a coherent manner.  This could be done at reasonable computing
expense because only the trivial zero-acceleration case would have to
be addressed (once the Earth's rotation is accounted for).  However,
since flux densities for pulsars in 47~Tuc certainly vary significantly
due to scintillation on time-scales of hours or less, the gain in
sensitivity might be quenched by periods within the 4.66\,h during
which we would be adding mostly noise to the time series being
analyzed.  In extreme cases, the sensitivity could actually be
degraded!  Optimally then, each session's data stream should be
analyzed with a multiplicity of integration spans.  This can, however,
degenerate into a huge overall data reduction task.  These
considerations also apply to the case of binary pulsars, where the
additional need to perform sensitive acceleration searches over longer
time intervals further exacerbates the overall computing load.

Finally, as already noted, we have analyzed most of the data at only
one value of DM.  The DMs for known pulsars in 47~Tuc span about one
unit (Table~\ref{tab:spin}), and this already introduces some loss of
signal for the shortest period pulsars we have considered
(\S\ref{sec:sens}).  It would also be advantageous to broaden the range
of DM searches because greater variance in the value of measured DMs
might add to the understanding of the properties of the ionized
intra-cluster, and interstellar, medium.  Most of the variance observed
for DMs of pulsars in globular clusters is expected to be due to
slightly different lines-of-sight across the Galaxy, because most
clusters are believed to be evacuated of gas (\cite{spe91}).

\subsection{Isolated Pulsars in 47~Tuc}\label{sec:population}

Mindful of the comments in the previous section, we can still make some
general statements about the likely intrinsic population of pulsars in
47~Tuc.  The observed population of pulsars, all with millisecond
periods, divides at present roughly into one-third isolated pulsars,
one-third normal binary systems ($P_b \sim 0.4$--2.3\,d and $m_2 \sim
0.2$\,M$_\odot$), and one-third short-period binary systems ($P_b \sim
0.06$--0.2\,d and $m_2 \sim 0.03$\,M$_\odot$).

As mentioned previously, the fraction of isolated pulsars known in
47~Tuc prior to this survey was 64\% of the total.  Use of acceleration
code has improved the detectability of binary systems, but we are still
selecting severely against binaries of the short-period variety, when
compared to isolated pulsars, and we consider it probable that the
actual fraction of isolated pulsars in 47~Tuc is less than the 35\%
found at present.  By comparison, the fraction of isolated millisecond
pulsars in the disk of the Galaxy is $\sim 20\%$ (e.g., \cite{cam98}).
The surveys of the Galactic disk, because of their short integration
times, are relatively more sensitive to short-period binaries, so the
isolated fraction for disk pulsars, up to small-number statistics, is
already known with some confidence.

The overall small fraction of isolated millisecond pulsars in 47~Tuc
suggests an interesting puzzle: long-period binaries ($P_b \sim
10$--$10^3$\,d) are expected to be formed at high rates in dense
clusters through exchange interactions between old isolated neutron
stars and ``hard'' primordial binaries (\cite{hmg+92}; \cite{sp93}).
The long-period binaries have, however, high cross-section to
disruption, and should eventually leave behind isolated millisecond
pulsars.  We detect no long-period binaries, and relatively few
isolated pulsars.  Perhaps most wide binaries are disrupted before the
neutron star companions can evolve into red giants, followed by mass
transfer and birth of millisecond pulsars, as in the standard scenario
(e.g., \cite{ver93}).  This is an open question.  In any case, even if
the isolated fraction of millisecond pulsars in 47~Tuc and in the
Galactic disk are comparable it is surely a coincidence, given the
different evolutionary pressures present in both environments.

\subsection{Profile Morphology}\label{sec:morph}

Chen \& Ruderman~(1993)\nocite{cr93} suggested that the formation
mechanism for millisecond pulsars in the Galactic disk and in globular
clusters might be different.  The observational evidence to support
this idea relied on supposed differences in the observed radio beams
(and hence magnetic field geometry, and origin) when comparing the
collection of pulse profiles then available for disk millisecond
pulsars and those in 47~Tuc, and in particular the supposed dearth of
interpulses among the pulsars in 47~Tuc.  More recently, Jayawardhana \&
Grindlay~(1996)\nocite{jg96} also presented evidence suggesting that
the observed pulse width distribution of disk and cluster millisecond
pulsars may be different.

Our collection of profiles for pulsars in 47~Tuc, shown in
Figure~\ref{fig:profs}, is larger and of higher quality than that
presented by Manchester et~al.~(1991)\nocite{mlr+91}, on which Chen \&
Ruderman~(1993)\nocite{cr93} and Jayawardhana \&
Grindlay~(1996)\nocite{jg96} based their conclusions.  In addition,
there are now many more high-quality pulse profiles available for disk
millisecond pulsars (\cite{kxl+98}; \cite{stc99}).  Given these
observational improvements, it is appropriate to briefly review the
situation regarding possible morphological differences.  In an attempt
to ensure homogeneity, we choose to compare our 20\,cm profiles with
the sample of 19 Galactic disk pulsars reported by Kramer et~al.~(1998)
which were also obtained from 20\,cm observations.

\medskip
\epsfxsize=8truecm
\epsfbox{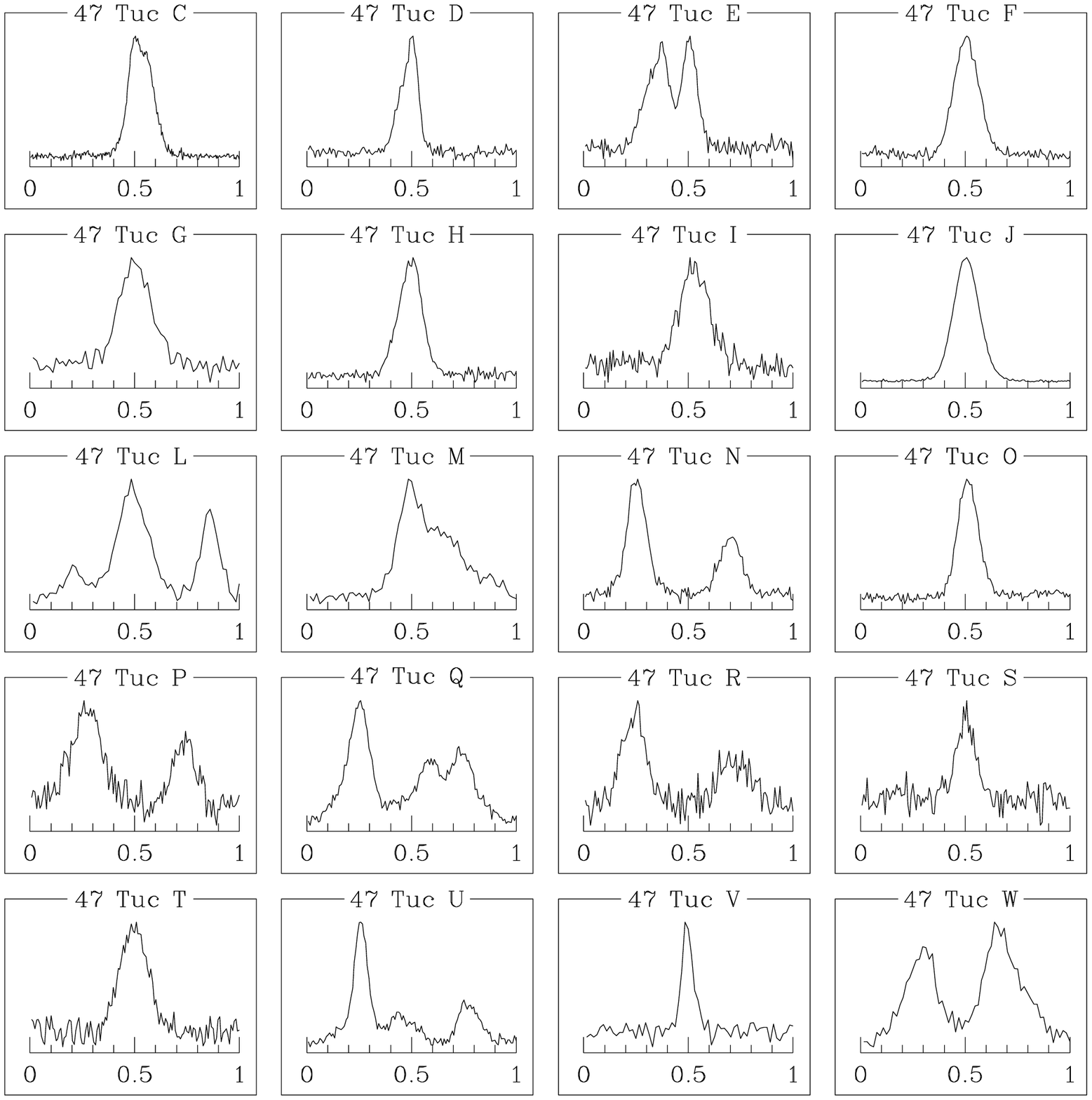}
\figcaption[profs.eps]{\label{fig:profs} Average pulse profiles at
20\,cm for pulsars in 47~Tucanae as a function of rotational phase.
These profiles are available in digital form at the European Pulsar
Network data archive
(\protect\verb+http://www.mpifr-bonn.mpg.de/div/pulsar/data+).  The
effective time resolution for all profiles displayed is $\sim 0.3$\,ms,
ranging from 0.04 in pulse phase for 47~Tuc~T to 0.14 for 47~Tuc~J.
Some profiles are not resolved (see Table~\protect\ref{tab:spin}).}
\bigskip

In order to test the suggestion that the pulse widths of the pulsars in
47~Tuc are broader than their counterparts in the Galactic disk, we
compared the pulse width distribution measured at the 50\% intensity
level ($w_{50}$) for the 20 pulsars listed in Table~\ref{tab:spin} with
that of the 19 disk millisecond pulsars presented in Table 2 of Kramer
et~al.~(1998).  A Kolmogorov-Smirnoff test returns a 25\% probability
that the two samples are drawn from the same underlying distribution.
At best this is a marginally significant result, reflecting the larger
fraction of Galactic disk millisecond pulsars for which $w_{50}<10$\%
(but note that in spite of the improved time resolution of the new
data, not all the profiles shown in Figure~\ref{fig:profs} are fully
resolved).

Finally, from Figure~\ref{fig:profs} we note that 11 pulsars have
essentially simple, single-component profiles; five pulsars have clear
interpulses; and four more have profiles that may be described as
complex, with emission evident across much of the period or in several
distinct components.  Inspecting the sample of Galactic disk
millisecond pulsars presented by Kramer et~al.~(1998),\nocite{kxl+98}
we find five clear interpulses from a sample of 24 pulsars.  Given the
present statistics, this is entirely consistent with the fraction of
interpulses we observe in 47~Tuc, refuting the morphological
differences suggested by Chen \& Ruderman~(1993)\nocite{cr93}.

\subsection{Eclipsing Pulsars}\label{sec:ecl}

Robinson et~al.~(1995)\nocite{rlm+95} show that, at 70 and 50\,cm,
47~Tuc~J is eclipsed by its companion for about a quarter of its
orbit.  We find that at 20\,cm, on the other hand, the pulsar is always
visible at all orbital phases.  However 47~Tuc~O, a pulsar system in
many respects similar to 47~Tuc~J, is always eclipsed for about
10--15\% of its orbit at 20\,cm.

There are three other similar systems known in 47~Tuc, with orbital
periods of less than 6\,h and with $\sim 0.03$\,M$_\odot$ companions.
47~Tuc~P and R were detected at only one epoch, although at least one
full orbit was covered in each case: 47~Tuc~P was seen at all orbital
phases, while 47~Tuc~R was apparently eclipsed for $\sim 25\%$ of its
orbit, centered with the pulsar at superior conjunction
(Fig.~\ref{fig:vel}).  47~Tuc~I is detected much more often, and there
are no eclipses observed at 20\,cm.

Finally, 47~Tuc~V also displays apparent eclipses, but we cannot yet
characterize the system or its eclipses, which may be irregular
(\S\ref{sec:resnew}); and 47~Tuc~W, with $m_2\sim 0.15$\,M$_\odot$, was
eclipsed for $\sim25\%$ of its orbit on the only occasion it was
detected (Fig.~\ref{fig:vel}).  While based on orbital parameters we
suggest that the system may be somewhat similar to the eclipsing pulsar
B1744$-$24A in Terzan~5 (\S\ref{sec:resnew}), we cannot yet be sure.

In the Galactic disk there are two pulsar systems known that have
properties broadly similar to some of the above eclipsing pulsars.
PSRs~B1957+20 and J2051$-$0827 (\cite{aft94}; \cite{sbl+96}) are
millisecond pulsars in very circular orbits with $\sim 0.03$\,M$_\odot$
companions, with orbital periods of a few hours.  These pulsars are
both eclipsed by their companions, apparently bloated to fill their
Roche lobes, but the systems differ in eclipse phenomenology:
PSR~B1957+20 always displays eclipses for $\sim 20\%$ of its orbit, at
all frequencies, while PSR~J2051$-$0827 is eclipsed at low radio
frequencies, but is visible at all orbital phases at 20\,cm; in the
latter instance, however, the radio pulses suffer extra time delays as
they propagate in a dispersive medium.  The differences observed in
these systems are likely due to a combination of different masses and
chemical compositions of the companions, and possibly evolutionary
states; the energy that is deposited at the companion by the
relativistic pulsar wind, which in turn depends on the pulsar spin
parameters, beaming geometry, and orbital separation; and geometrical
effects.  It is not yet clear if the eclipse behavior for 47~Tuc~J or
O, let alone that for the seldom-detected pulsars in 47~Tuc, is similar
to that of either PSR~J2051$-$0827 or B1957+20.

\subsection{The Central Mass Density of 47~Tuc}\label{sec:density}

In Table~\ref{tab:3sol} we list phase-connected timing solutions for
three pulsars.  An unusual feature of all the solutions is that the
period derivatives, $\dot P$, are negative, apparently implying the
pulsars are spinning up, rather than spinning down as is the case for
all pulsars known in the Galactic disk, which are losing rotational
energy emitted in the form of magnetic dipole radiation and a
relativistic particle wind.  In fact this is a feature observed in some
other globular cluster pulsars (see, e.g., \cite{wkm+89}), and is most
easily interpreted by considering that the observed $\dot P_{\rm obs}$
is a combination of the intrinsic $\dot P_{\rm int} (\ge0)$ and a
contribution resulting from the acceleration of the pulsar in the
gravitational potential of the cluster, such that
\begin{equation}
\label{eq:pdot}
\dot P_{\rm obs} = \dot P_{\rm int} + \dot P_{\rm acc}.
\end{equation}
In this equation, $\dot P_{\rm acc}/P = a_l/c$, where $a_l$ is the
acceleration along the line of sight (\cite{phi92b}).  A negative $\dot
P_{\rm obs}$ implies the pulsar lies behind the cluster center as seen
from the Sun.

Given a model for the mass distribution in the cluster, a lower limit
on $a_l$ by assuming $\dot P_{\rm int} = 0$, and an accurate pulsar
position, one can estimate a lower limit to the central mass density in
the cluster.  Assuming that the central regions of the cluster can be
described by King's (1962) \nocite{kin62} empirical density model, one
obtains for the central central mass density
\begin{equation}
\label{eq:rho}
\rho(0) > \left| \frac{a_l r}{4 \pi G \kappa^2} \right| = \left|
\frac{9 c r \dot P} {4 \pi G r_{c}^2 P} \right|.
\end{equation}
In this expression $r$ is the projected distance of the pulsar from the
cluster center and $\kappa$ is the Jeans length at the cluster center.
For a King model, we have made the standard approximation that $\kappa$
is one third of the core radius of the cluster, $r_c$.  We use
equation~\ref{eq:rho} for the pulsars in 47~Tuc by taking the most
recent and reliable determination of the angular core radius, based on
a fit of a King radial density profile to the observed stellar
population; together with a distance estimate to
47~Tuc (Table~\ref{tab:47tuc}), equation~\ref{eq:rho} becomes
\begin{equation}
\label{eq:rho2}
\rho(0) > 2.8\times10^5\,\left(\frac{R}{\rm arcmin}\right)
\times\left(\frac{\dot P_{\rm acc}}{10^{-20}}\right)
\times\left(\frac{P}{\rm ms}\right)^{-1}\,{\rm M}_{\odot}\,{\rm pc}^{-3}.
\end{equation}
Here $R$ is the angular separation between the pulsar and the center of
the cluster (see Table \ref{tab:47tuc}).  Using the celestial
coordinates, $P$, and $\dot P$, listed in Table~\ref{tab:3sol} for
47~Tuc~C, D, and J, we obtain from equation~\ref{eq:rho2} lower limits
on $\rho(0)$ of, respectively, 30, 1, and
$13\times10^4$\,M$_\odot\,{\rm pc}^{-3}$.

The most stringent lower limit on $\rho(0)$, which we obtain from the
position and period derivative of 47~Tuc~C, is some four times larger
than the value quoted by Robinson et~al.~(1995)\nocite{rlm+95} for the
same pulsar.  This is due to our use of the latest core radius estimate
(\cite{dps+96}) which is approximately half the value of previous
estimates (see, e.g., \cite{cdps93}).  De Marchi et~al.  demonstrate
that the difference in core radius estimates can be entirely attributed
to the effects of photometric incompleteness in the earlier samples.

If a King model can be used to describe the central regions of the
cluster, and the observations of De Marchi et~al.~(1996) suggest that
this is the case, the above are hard limits on $\rho(0)$, due to the
assumption so far that $\dot P_{\rm int} = 0$.  In fact $\dot P_{\rm
int}$ is likely to have some finite positive value, such that we have
thus far underestimated the accelerations felt by the pulsars.  We now
present some plausibility arguments that lead in general to more
realistic estimates of a limit on $\rho(0)$.

Most millisecond pulsars known in the Galactic disk ($1.5<P<9$\,ms)
have inferred intrinsic surface dipole magnetic field strengths of
$10^8$--$10^9$\,G, with an average of $B \sim 3\times10^8$\,G (e.g.,
\cite{ctk94}).  For a given $P$ and $B$, the period derivative is
obtained (e.g., \cite{mt77}) from
\begin{equation}
\label{eq:b}
\dot P = 10^{-19}\left(\frac{P}{\rm ms}\right)^{-1}\times
\left( \frac{B}{3.2\times10^8\,{\rm G}} \right)^2.
\end{equation}

We have no a priori information about the magnetic field strengths of
the pulsars in 47~Tuc, but if they were significantly higher than those
for disk pulsars, so would the respective $\dot P$.  For a putative
6\,ms pulsar, like 47~Tuc~C, but with $B=10^{10}$\,G, we obtain $\dot
P=2\times10^{-17}$ from equation~\ref{eq:b}.  In turn, when used with
equation~\ref{eq:rho2}, this would imply
$\rho(0)>3\times10^7$\,M$_\odot$\,pc$^{-3}$, an unreasonably large
value!  A similar argument can be made for 47~Tuc~J.  If the magnetic
fields of the millisecond pulsars in 47~Tuc are not likely to be much
larger than those of millisecond pulsars in the disk, we may assume
that they are, on average, comparable.

Following the above arguments, we calculate from equation~\ref{eq:b}
the likely values of $\dot P_{\rm int}$ for 47~Tuc~C and J:
$1.7\times10^{-20}$ and $4.8\times10^{-20}$ respectively, for
$B=3.2\times10^8$\,G.  These expected values can be decreased or
increased by a factor of ten, considering the entire range
$10^8<B<10^9$\,G.  To be conservative, we consider only
$10^8<B<3.2\times10^8$\,G.  For 47~Tuc~C, the full contribution due to
acceleration in the cluster (from eq.~\ref{eq:pdot}), is
5.1--6.7$\times10^{-20}$, while for 47~Tuc~J it is
1.4--5.7$\times10^{-20}$.  These contributions are increased over the
hard limits assuming $\dot P_{\rm int} = 0$ by factors of 1.0--1.3 for
47~Tuc~C and 1.5--6.1 for 47~Tuc~J.  Finally, the implied lower limits
on $\rho(0)$ are increased to 3--4$\times10^5$\,M$_\odot$\,pc$^{-3}$
and 2--10$\times10^5$\,M$_\odot$\,pc$^{-3}$, for 47~Tuc~C and J
respectively.

We conclude from the present timing solutions that
$\rho(0)>3\times10^5$\,M$_\odot$\,pc$^{-3}$.  This lower limit can be
compared to ``standard estimations'' of $\rho(0)$, based on optical
measurements of King model parameters (e.g., \cite{web85}).  The
aforementioned revision in the core radius value for 47~Tuc implies
that the existing estimate for $\rho(0)$, which scales as $r_c^{-2}$
(\cite{kin66}; \cite{spi87}), should be revised appropriately.  Scaling
the value for $\rho(0)$ tabulated by \nocite{web85} Webbink (1985), we
find $\rho(0)=4\times10^5$\,M$_\odot\,{\rm pc}^{-3}$.  While this is in
agreement with our present lower limits, it is unlikely that both
47~Tuc~C and J have the lowest magnetic field strengths of any known
pulsars and so, in the context of a King model, the above discussion is
strongly suggestive that for 47~Tuc,
$\rho(0)>4\times10^5$\,M$_\odot$\,pc$^{-3}$, and possibly is as much as
a factor of two or so larger.

\subsection{Pulsar Luminosities}\label{sec:flux}

Except for 47~Tuc~C, D, and J, pulsars in 47~Tuc are detected on fewer
than 75\% of observing days, and we have detected some of the pulsars
on only one of 75 observing days (see Tables~\ref{tab:obs} and
\ref{tab:rare}).  Given this fact, it is not appropriate to estimate
average flux densities for most pulsars by simply averaging the
detected signal-to-noise ratios and converting to flux densities using
the approximately known system noise and detection apparatus
characteristics.  In this section we present a more rigorous method to
estimate mean flux densities of the pulsars in 47~Tuc.

The variability in apparent flux densities, apart from the occurrence
of eclipses, results from interstellar scintillation (\cite{ric77}),
and we proceed further by assuming that scintillation affects all
pulsars in 47~Tuc similarly, since they are all located in relatively
close proximity.

Given this assumption, we construct cumulative histograms in
$\log(\sigma)$ for the frequently-detected pulsars in the cluster, and
attempt to synthesize a ``standard'' histogram which we take to be
representative of the relative flux variations due to interstellar
scintillation for a pulsar in 47~Tuc.  Initially the standard histogram
is defined to be that of 47~Tuc~J, the brightest pulsar.  The
histograms of other frequently-appearing pulsars are then shifted by a
scaling factor in a fitting procedure so as to match the standard as
closely as possible, in a least-squares sense.  A new standard is then
defined where $\sigma$ of the $n^{\rm th}$ strongest ``observation'' is
the average of the respective values from each of the individually
shifted histograms, and this procedure is iterated until the scaling
factors are stationary and a self-consistent solution is obtained.  The
scaling factors with respect to the final standard (see
Fig.~\ref{fig:shist}) therefore provide an estimate of the average
relative signal-to-noise ratios of the various pulsars.

\medskip
\epsfxsize=8truecm
\epsfbox{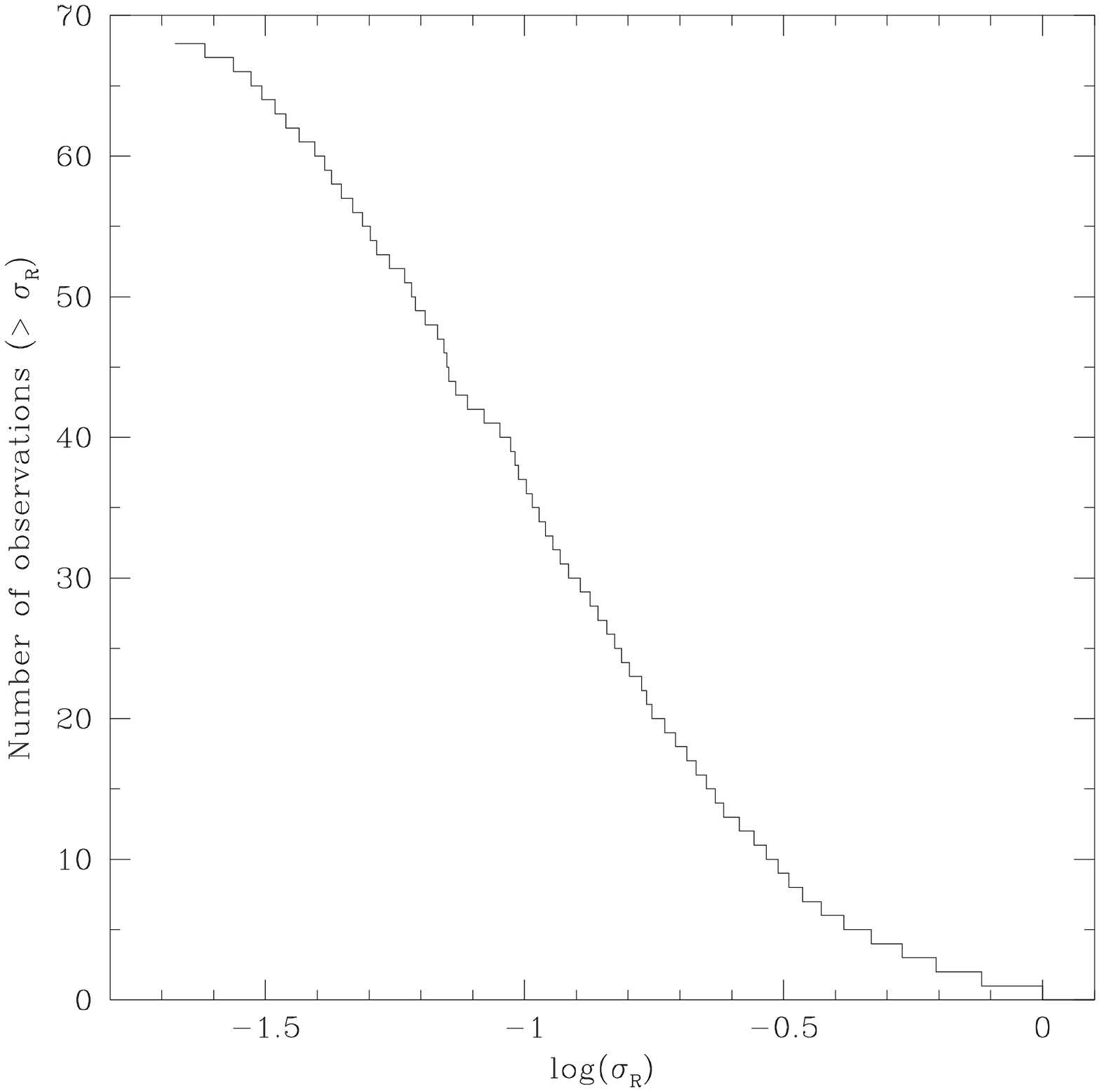}
\figcaption[shist.eps]{\label{fig:shist} Cumulative ``standard''
histogram of $\log(\sigma_R)$ for observations of pulsars in 47~Tuc.
$\sigma_R$ is the signal-to-noise ratio of an ``observation'',
$\sigma$, scaled to the maximum signal-to-noise ratio, $\sigma_{\rm
max}$, i.e., $\sigma_R=\sigma/\sigma_{\rm max}$.  See
\protect\S\protect\ref{sec:flux} for details.}
\bigskip

For 47~Tuc~J we obtain the actual average $\sigma$ from the
observations (Table~\ref{tab:obs}), but correct it downwards to take
into account the $\sim 10\%$ of observing days in which we do not
detect the pulsar.  We do this by assuming that on these days, on
average, the pulsar has a flux density that is half of the lowest value
detected, resulting in a 10\% correction.  After applying the scaling
factors and converting to a flux density scale using the methodology
outlined in \S\ref{sec:sens}, we obtain the average flux densities
listed in Table~\ref{tab:spin} for 14 pulsars.

One obvious problem with this method is that for the less-frequently
appearing pulsars only part of the cumulative histogram is being fitted
to --- and to the extent that the standard histogram is in fact not
precisely descriptive of the scintillation behavior of all pulsars,
this may particularly skew the average $\sigma$ for such pulsars.  We
have accounted for this in the uncertainties listed in
Table~\ref{tab:spin}.  In addition, until we calibrate the sensitivity
of the observing system with a pulsed source of known strength, the
conversion between $\sigma$ and flux density is subject to a systematic
uncertainty that we judge to be of the order of $\sim 25\%$.

Pulsars 47~Tuc~C, D, and J, for which we have the most reliable flux
density estimates, display maximum ``daily'' flux densities that are
factors of 4--8 greater than the average values.  This is the only
information we have available that is of some use to estimate, very
roughly, the average flux density for the pulsars that are detected
very infrequently (Table~\ref{tab:rare}).  We obtain estimates of $\sim
0.04$\,mJy for the average flux density of these pulsars.  We note this
plausible estimate simply to suggest that the flux densities of these
pulsars need not be much lower than those of the weakest among the more
frequently-appearing pulsars.

The weakest pulsars we have been able to detect in 47~Tuc have flux
densities at 1400\,MHz of about $S_{1400} = 0.04$\,mJy, while the
strongest, 47~Tuc~J, has $S_{1400} = 0.54$\,mJy.  At the distance of
47~Tuc (Table~\ref{tab:47tuc}), these correspond to luminosities at
1400\,MHz, $L_{1400} \equiv S_{1400} d^2$, of 0.8--11\,mJy\,kpc$^2$.
Translated to luminosities at frequencies $\sim 400$\,MHz, assuming a
spectral index for the pulsars in 47~Tuc of $-2.0$, these correspond
roughly to $10 < L_{400} < 130$\,mJy\,kpc$^2$.

The limiting minimum luminosity for Galactic disk pulsars at $\sim
400$\,MHz is about 1\,mJy\,kpc$^2$ or lower (\cite{lbdh93};
\cite{lml+98}).  For the Galactic disk (\cite{lmt85}) and M15
(\cite{and92}), the differential luminosity distribution for pulsars is
found to be consistent with $dN = L^{-1}\,d\log L$.  Assuming that
these findings apply to 47~Tuc, we estimate there are about 10 times as
many potentially observable pulsars in the luminosity decade
1--10\,mJy\,kpc$^2$ as there are in 10--100\,mJy\,kpc$^2$, where we
already know of about 20 pulsars.  Hence we estimate a potentially
observable population of pulsars in 47~Tuc, similar to the ones already
detected, of about 200, above a minimum luminosity $L_{400} \sim
1$\,mJy\,kpc$^2$.  When beaming and selection effects are accounted for
we judge it likely that 47~Tuc contains several hundred pulsars with
$L_{400} \ga 1\,$mJy\,kpc$^2$.

The actual determination of the luminosity distribution, and a detailed
discussion of flux densities, spectral indices, and scintillation
properties for the pulsars in 47~Tuc, will rely on more complete
multi-wavelength data sets, and will be addressed elsewhere.  It is
already clear however, that we have detected only the few brightest
pulsars in 47~Tuc, and continued searches of this cluster, particularly
with greater sensitivity, should yield a bounty of new discoveries.

\acknowledgements

We thank the skilled and dedicated telescope staff at Parkes for their
support during this project, and Nichi D'Amico, Vicky Kaspi, Froney
Crawford, and Jon Bell for assistance with observations.  The Parkes
telescope is part of the Australia Telescope which is funded by the
Commonwealth of Australia for operation as a National Facility managed
by CSIRO.  Arecibo Observatory is operated by Cornell University under
cooperative agreement with the National Science Foundation.  F.C.
gratefully acknowledges support from the European Commission through a
Marie Curie fellowship under contract ERB~FMBI~CT961700, and the
gracious hospitality provided by Jules Halpern at Columbia University
during the completion of this work.  P.F. acknowledges support from the
Funda\c{c}\~{a}o para a Ci\^{e}ncia e a Tecnologia through a Praxis~XXI
fellowship under contract BD/11446/97.  We also thank Andy Fruchter and
Miller Goss for generously sharing some data prior to publication, and
we are grateful to Fred Rasio, David Nice, and the anonymous referee,
for useful comments on an earlier version of the manuscript.

\end{document}